\documentclass{article}
\usepackage{arxiv}

\usepackage{amsmath,amssymb}
\usepackage[utf8]{inputenc} 
\usepackage[T1]{fontenc}    
\usepackage{hyperref}       
\usepackage{url}            
\usepackage{booktabs}       
\usepackage{amsfonts}       
\usepackage{nicefrac}       
\usepackage{microtype}      
\usepackage{lipsum}
\usepackage{siunitx}
\usepackage[linesnumbered,ruled,vlined]{algorithm2e}
\usepackage{float}
\usepackage{amsmath}
\usepackage{amsthm}
\usepackage{mathtools}
\usepackage{enumerate}
\usepackage{cite}
\usepackage{xcolor}
\usepackage{graphicx}
\usepackage{braket}
\usepackage{comment}
\usepackage{caption}
\usepackage{cleveref}
\usepackage{subcaption}
\graphicspath{ {./images/} }
\newcommand{\tr}{\mathrm{tr}}

\DeclareMathOperator*{\argmin}{argmin}

\title{Weighted Approximate Quantum Natural Gradient for Variational Quantum Eigensolver}

\author{
  Chenyu Shi\\
  Applied Quantum Algorithms Leiden, Leiden University, Leiden, The Netherlands\\
  Leiden Institute of Advanced Computer Science, Leiden University, Leiden, The Netherlands\\
  \texttt{c.shi@liacs.leidenuniv.nl}\\
  \And
  Vedran Dunjko\\
  Applied Quantum Algorithms Leiden, Leiden University, Leiden, The Netherlands\\
  Leiden Institute of Advanced Computer Science, Leiden University, Leiden, The Netherlands\\
  \texttt{v.dunjko@liacs.leidenuniv.nl}\\  
  \And
  Hao Wang\thanks{Corresponding author}\\
  Applied Quantum Algorithms Leiden, Leiden University, Leiden, The Netherlands\\
  Leiden Institute of Advanced Computer Science, Leiden University, Leiden, The Netherlands\\
  \texttt{h.wang@liacs.leidenuniv.nl}\\
}

\begin{document}
\maketitle

\begin{abstract}
The variational quantum eigensolver (VQE) is one of the most prominent algorithms using near-term quantum devices, designed to find the ground state of a Hamiltonian. In VQE, a classical optimizer iteratively updates the parameters in the quantum circuit. Among various optimization methods, the quantum natural gradient descent (QNG) stands out as a promising optimization approach for VQE. However, standard QNG only leverages the quantum Fisher information of the entire system and treats each subsystem equally in the optimization process, without accounting for the different weights and contributions of each subsystem corresponding to each local term in the Hamiltonian. To address this limitation, we propose a Weighted Approximate Quantum Natural Gradient (WA-QNG) method tailored for $k$-local Hamiltonians. In this paper, we theoretically analyze the potential advantages of WA-QNG compared to QNG from three distinct perspectives and reveal its connection with the Gauss-Newton method. We also show it outperforms the standard quantum natural gradient descent in the numerical simulations for seeking the ground state of the Hamiltonian.
\end{abstract}

\section{Introduction}
Quantum computing is widely regarded as having potential advances in numerous fields. However, due to the limitations of the noise and scale of current Noisy Intermediate-Scale (NISQ) quantum devices \cite{preskill2018quantum}, algorithms such as Shor's \cite{shor1999polynomial} and Grover's \cite{grover1996fast} still remain beyond practical implementation. A computational paradigm well-suited for current NISQ quantum devices is that of the variational quantum algorithms, which is a kind of variational hybrid approach \cite{bharti2022noisy}. These algorithms leverage a feedback loop between classical and quantum computers. In this paradigm, the quantum computer evaluates an objective function formulated by a parameterized quantum circuit, while the classical computer employs an optimizer to iteratively update the circuit parameters to seek the optimal value \cite{moll2018quantum}.  

Variational quantum algorithms have drawn significant attention across various fields, including quantum physics and quantum chemistry \cite{aspuru2005simulated, peruzzo2014variational}, optimization \cite{farhi2014quantum, grange2023introduction}, and machine learning \cite{havlivcek2019supervised, jerbi2023quantum}. Among these, one of the most well-known variational quantum algorithms is the Variational Quantum Eigensolver (VQE) \cite{peruzzo2014variational}. VQE is designed to find the ground state of a given quantum system. In this algorithm, a variational quantum circuit is employed to estimate the expectation value of the system with respect to a given Hamiltonian. Additionally, a classical optimizer iteratively updates the parameters in the quantum circuit to minimize the expectation value. Through this optimization process, the algorithm is expected to converge to a solution that closely approximates the ground state. Despite the high potential in this field, VQEs also face several challenges during optimization, such as barren plateaus \cite{larocca2025barren, larocca2024review, sack2022avoiding, park2024hamiltonian}. Therefore, developing efficient optimization methods for VQEs is of great importance.

The optimizer in VQE plays a crucial role in determining the algorithm's performance. In addition to the most basic gradient descent method (referred to as vanilla gradient descent in this context), more advanced variants such as stochastic gradient descent (SGD) \cite{robbins1951stochastic} and Adam \cite{diederik2014adam} are widely adopted. Among these, the quantum natural gradient descent (QNG) \cite{stokes2020quantum} emerges as a promising approach. It is the quantum analog of natural gradient descent \cite{amari1998natural, martens2020new} in its classical counterpart. QNG leverages the quantum Fisher information matrix \cite{liu2020quantum} of the quantum system. It captures the geometric information and is expected to obtain better convergence performance in the optimization process.

In the standard formulation of QNG, the quantum Fisher information used in the optimization step corresponds to the entire quantum system. However, in VQE, particularly in quantum chemistry, the Hamiltonian $H$ is often expressed as a summation of several local terms $H_m$ with maximum locality $k$, where $H=\sum_m h_mH_m$ with the corresponding output of the quantum circuit $\tr(\rho H)=\sum_m h_m\tr(\rho_mH_m)$. Recent studies \cite{park2024hamiltonian, anuar2024operator} have shown that leveraging the locality structure of $k$-local Hamiltonian can improve the performance of finding ground state. 

For a $k$-local Hamiltonian, each subsystem $\rho_m$ corresponding to each local term $H_m$ contributes differently to the final output of the quantum circuit due to the different weights $h_m$. Therefore, a potential improvement for standard QNG is to assign these subsystems with different weights during the training, rather than only using the quantum Fisher information matrix of the entire system, where all subsystems are treated equally. Hence, here we propose a new approach, the Weighted Approximate Quantum Natural Gradient Descent (WA-QNG), which takes the different weights and contributions of the subsystem corresponding to each local observable term into account. 

In WA-QNG, we replace the quantum Fisher information matrix of the entire quantum system with the weighted summation of the Hilbert-Schmidt metric tensors of the subsystems corresponding to each local term in the optimization step. We theoretically analyze the potential advantages of WA-QNG compared to QNG from three distinct perspectives and reveal its connection to the Gauss-Newton method. Our method displays efficient convergence speed compared to standard QNG in the numerical simulations. Furthermore, we show that the Hilbert-Schmidt metric tensor required for WA-QNG can be efficiently estimated using the classical shadow method \cite{huang2020predicting} for $k$-local Hamiltonians.

The remainder of the paper will be structured as follows. \cref{back} introduces the preliminary background knowledge, including the variational quantum eigensolver and the quantum natural gradient descent. \cref{waqng} formulates the WA-QNG method and theoretically analyzes its potential advantages over standard QNG from three different perspectives. \cref{OI} explores the connection between WA-QNG and the Gauss-Newton method. \cref{CAS} discusses the computational costs of WA-QNG compared to standard QNG. \cref{exp} presents the numerical results to support our theoretical analysis. Finally, \cref{con} concludes the paper.

\section{Background}\label{back}
In this section, we provide a brief overview of the foundational background relevant to this paper. First, we briefly introduce VQE and explain its working principles. Then, we give the definition of the quantum natural gradient descent and discuss its relation with quantum Fisher information.
\subsection{Variational Quantum Eigensolver} \label{vqe}
\begin{figure*}[ht]
  \centering
  \includegraphics[width=0.8\textwidth]{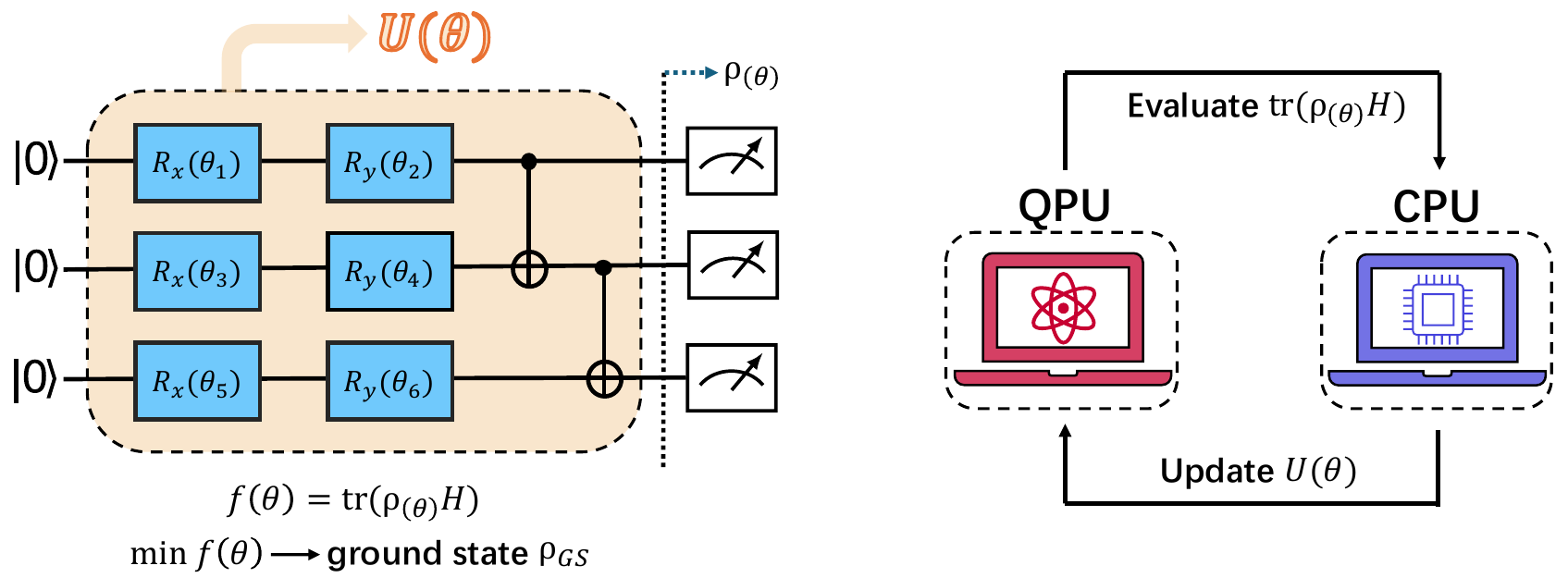}
  \caption{Left: An illustration of the variational quantum circuit in VQE. A parameterized quantum circuit $U(\theta)$ is used to prepare a variational quantum state $\rho_{\theta}$. By adjusting the parameter $\theta$, the quantum circuit aims to prepare the state $\rho_{\theta}$ that approximates the ground state $\rho_{GS}$ by minimizing the objective function $f(\theta)$. Right: The general working principle of VQE. The quantum computer evaluates the expectation value $f(\theta)=\tr(\rho_{\theta}H)$, while the classical computer employs an optimizer to iteratively update the parameters $\theta$ to minimize the objective function $f(\theta)$.}
  \label{fig:vqe}
\end{figure*}

The goal of the VQE, initially introduced by \cite{peruzzo2014variational}, is to approximately seek the ground state $\rho_{GS}$ with respect to a Hamiltonian $H$. A variational quantum circuit is used to prepare a variational quantum state $\rho_{\theta}$, where $\rho_{\theta}=U(\theta)\ket{0}\bra{0}U^{\dag}(\theta)$. The expectation value with respect to the Hamiltonian $H$ is evaluated by the quantum computer. An illustration of variational quantum circuit in VQE is given in \cref{fig:vqe}. Hence, the variation quantum circuit realizes the following function $f$:
\begin{equation}
f(\theta) = \tr(\rho_{\theta}H)
\label{eq:vqe}
\end{equation}

According to the definition, the ground state $\rho_{GS}$ is the lowest energy state of the given Hamiltonian $H$. Hence, seeking the ground state by adjusting the parameter $\theta$ is equivalent to minimizing the function $f(\theta)$.

The value of function $f(\theta)$ is fed to a classical computer in the optimization process. The classical computer employs an optimizer, where $f(\theta)$ is the objective function, to iteratively update the parameters in the quantum circuit. The most common optimization method is the vanilla gradient descent. In each optimization step, the parameters are updated by:
\begin{equation}
\theta^{(k+1)}=\theta^{(k)}-\eta \nabla f(\theta^{(k)})
\label{eq:vanilla}
\end{equation}

where $\nabla f$ is the gradient of the objective function and $\eta$ is the learning rate. In general, the gradient can be approximated using a naive finite difference method. In VQE, however, the most common approach to obtain the exact gradient is through the parameter-shift rule \cite{mitarai2018quantum, schuld2019evaluating}, up to finite sampling errors.

The general working principle of VQE is also illustrated in  \cref{fig:vqe}. After sufficient training with an expressive variational circuit, VQE is expected to produce a good approximation of the ground state.

\subsection{Quantum Natural Gradient Descent} \label{qngd}
Each optimization update step in vanilla gradient descent can be formulated as the following constrained optimization problem \cite{ollivier2017information,martens2020new} for a small $\varepsilon$ with a small change $\delta$ in parameter space:

\begin{equation}\label{vanilla_problem}
\begin{array}{cl}
\min\limits_{\delta} &\; f(\theta + \delta) \\
\mbox{s.t.} &\; \Vert \delta \Vert^2_2 \leq \varepsilon
\end{array}
\end{equation}

This constrained optimization problem seeks to minimize the objective function $f$ within a local neighborhood of $\theta^{(k)}$. Note that solving this problem by applying a first-order Taylor approximation to $f$ leads to the derivation of the vanilla gradient descent step in ~\cref{eq:vanilla}.

A limitation of this method is that each step is inherently tied to the Euclidean geometry of the parameter space, as the Euclidean distance is used to define the local neighborhood in the constrained optimization problem. However, the distances in the optimization landscape can be distorted in reparameterization -- for example, directions that were equally steep may become scaled differently, potentially leading to inefficiencies in the optimization process \cite{amari1998natural, stokes2020quantum}.

\begin{figure*}[ht]
  \centering
  \includegraphics[width=0.65\textwidth]{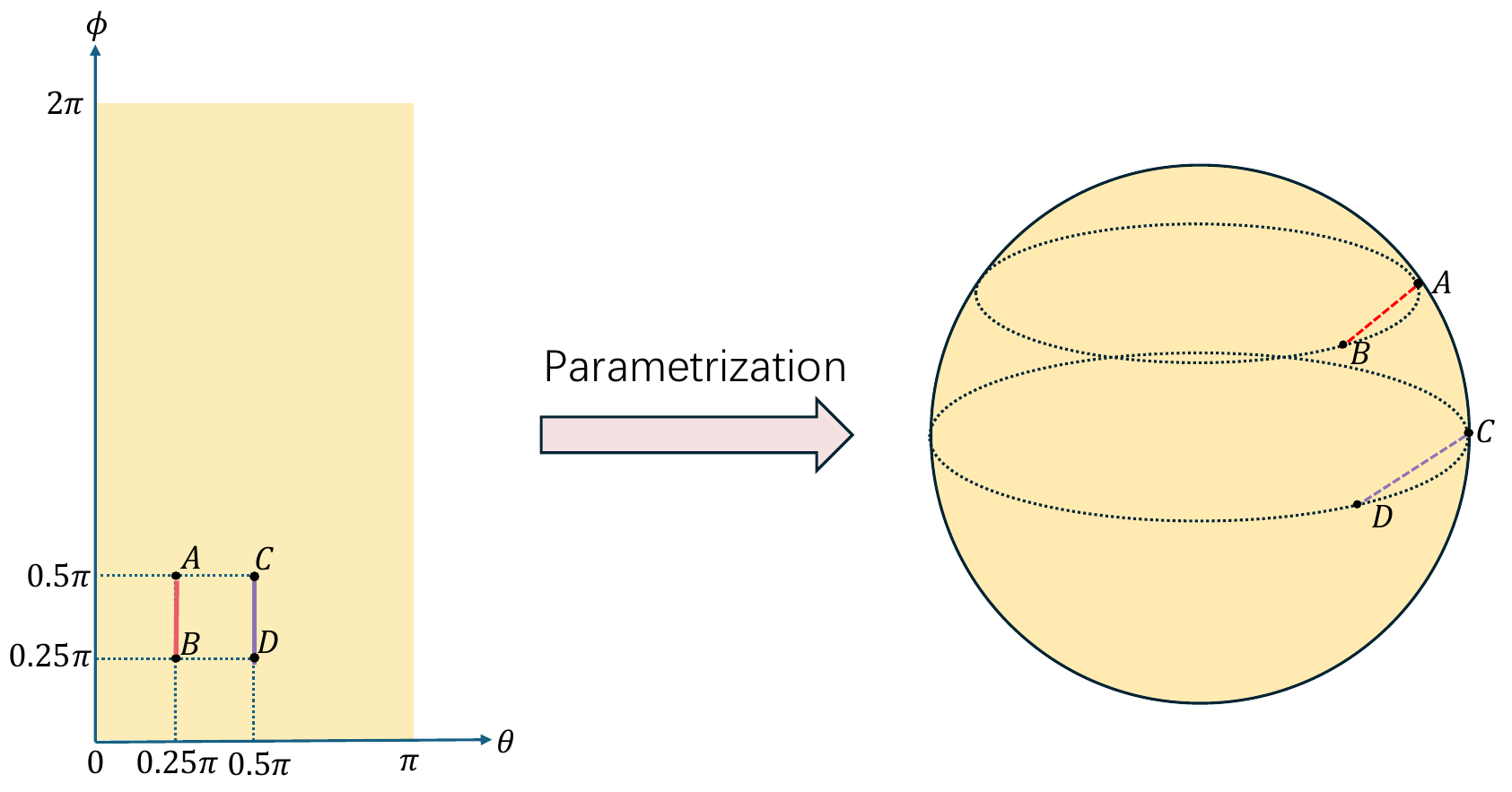}
  \caption{An illustration of distance distortion in the parametrization. The original parameter space is an Euclidean space. The parametrization maps the point in the parameter space to a point on the sphere with radius $r=1$ by the coordinate transformation $x=\sin(\theta)\cos(\phi)$, $y=\sin(\theta)\sin(\phi)$, $z=\cos(\theta)$. The distances from $A$ to $B$ (red line) and from $C$ to $D$ (purple line) are the same in the original parameter space. However, after parametrization to the sphere, the distances are distorted and become different.}
  \label{fig:param}
\end{figure*}

An illustration is shown in \cref{fig:param}. The parameter space is a Euclidean space $[0, \pi] \times [0, 2\pi]$, where each coordinate corresponds to the polar angle and azimuthal angle of a sphere. The parametrization maps each point in the original parameter space to a point on the surface of a unit sphere using the coordinate transformation $x=\sin(\theta)\cos(\phi)$, $y=\sin(\theta)\sin(\phi)$, $z=\cos(\theta)$. In the original parameter space, the distances between points $A$ and $B$ (red line) and between points $C$ and $D$ (purple line) are the same in terms of Euclidean distance. However, after parametrization onto the sphere, their distances differ significantly \footnote{In \cref{fig:param}, we use the chord length as the distance metric between two points on the sphere for simplicity. Note that the chord length is always proportional to the great-circle distance, which represents the true geodesic distance on the sphere.}. An intuitive example is as follows: Suppose the original parameter space consists of latitude and longitude pairs, and the parametrization maps each pair to a point on the surface of the Earth. It is more natural to describe distances using the great-circle path distance defined directly on the Earth's surface (i.e., distance after parametrization) rather than the Euclidean distance between latitude and longitude pairs (i.e., distance in the parameter space).

Similarly, in the parametrization from the parameter $\theta$ to the variational quantum state $\rho_{\theta}$, the distance distortion can also occur. Changes with the same Euclidean distance in the parameter space can result in different changes in the variational quantum state $\rho_{\theta}$. Therefore, it is more natural to use a distance metric defined directly for the variational quantum state $\rho_{\theta}$ to reformulate the constrained optimization problem, rather than relying on the Euclidean distance defined for $\theta$ in the original parameter space:
\begin{equation}\label{qng_problem}
\begin{array}{cl}
\min\limits_{\delta} &\; f(\theta + \delta) \\
\mbox{s.t.} &\; D_{F}(\rho_{\theta}, \rho_{\theta +\delta}) \leq \varepsilon
\end{array}
\end{equation}
where the distance metric $D_F(\rho, \sigma) = 1 - (\tr(\sqrt{\sqrt{\rho}\sigma \sqrt{\rho}}))^2$ is the fidelity distance. For pure states $\rho=\ket{\phi}\bra{\phi}$ and $\sigma=\ket{\psi}\bra{\psi}$, the fidelity distance can also be formulated as $D_F(\ket{\phi}, \ket{\psi})=1-\vert \braket{\phi | \psi}\vert^2$. Solving the above optimization problem derives the quantum natural gradient (QNG) update step:

\begin{equation}
\theta^{(k+1)}=\theta^{(k)}-\eta F^{+} \nabla f(\theta^{(k)})
\label{eq:qng}
\end{equation}

where $F^{+}$ is the pseudo-inverse of the quantum Fisher information matrix $F$ at $\theta^{(k)}$, and $\eta$ is the learning rate. For a pure state $\ket{\psi(\theta)}$, the quantum Fisher information matrix $F$ is given by:
\begin{equation}
F_{ij} = 4\operatorname{Re} \Big \{\braket{\frac{\partial \psi}{\partial \theta_i}|\frac{\partial \psi}{\partial \theta_j}}-\braket{\frac{\partial \psi}{\partial \theta_i}|\psi}\braket{\psi|\frac{\partial \psi}{\partial \theta_j}} \Big \}
\label{eq:fisher}
\end{equation}

where $\theta_i$ denotes the $i$-th element of the parameter $\theta$, and $F_{ij}$ represents the $(i,j)$-th entry of the quantum Fisher information matrix $F$. For details on the derivation, please refer to \cite{stokes2020quantum} and \cite{PhysRevA.105.052402}. A detailed discussion of the derivation is also provided in \cref{GEO}. QNG has been shown to achieve better performance compared to vanilla gradient descent in previous studies \cite{stokes2020quantum, amari1998natural, yamamoto2019natural}.

\section{WA-QNG: Weighted Approximate Quantum Natural Gradient Descent}\label{waqng}
In this section, we present the formulation of the Weighted Approximate Quantum Natural Gradient Descent (WA-QNG) method and discuss its potential advantages. First, we highlight the limitation of QNG, where the weights of subsystems corresponding to each local term are not considered. To address this issue, we introduce WA-QNG, which leverages the weighted sum of the Hilbert-Schmidt metric tensors of each subsystem in the optimization step. Additionally, we demonstrate WA-QNG's potential advantages from three different perspectives. 

\subsection{Limitation of QNG} \label{LQ}
The quantum Fisher information matrix of $\rho_{\theta}$ in~\cref{eq:qng} represents the quantum Fisher information matrix of the entire quantum system. A key limitation is that this quantum Fisher information does not incorporate any information about the Hamiltonian. Consequently, QNG utilizes the same quantum Fisher information matrix $F$ in the update formula for two different Hamiltonians, which captures the sensitivity of the quantum state with respect to parameter changes. However, in VQE, we care more about how the sensitivity of the final objective function value with respect to parameter changes, rather than the sensitivity of the quantum state with respect to parameter changes itself. Therefore, information about the Hamiltonian should also be taken into account during the optimization process.

\begin{figure*}[ht]
  \centering
  \includegraphics[width=0.5\textwidth]{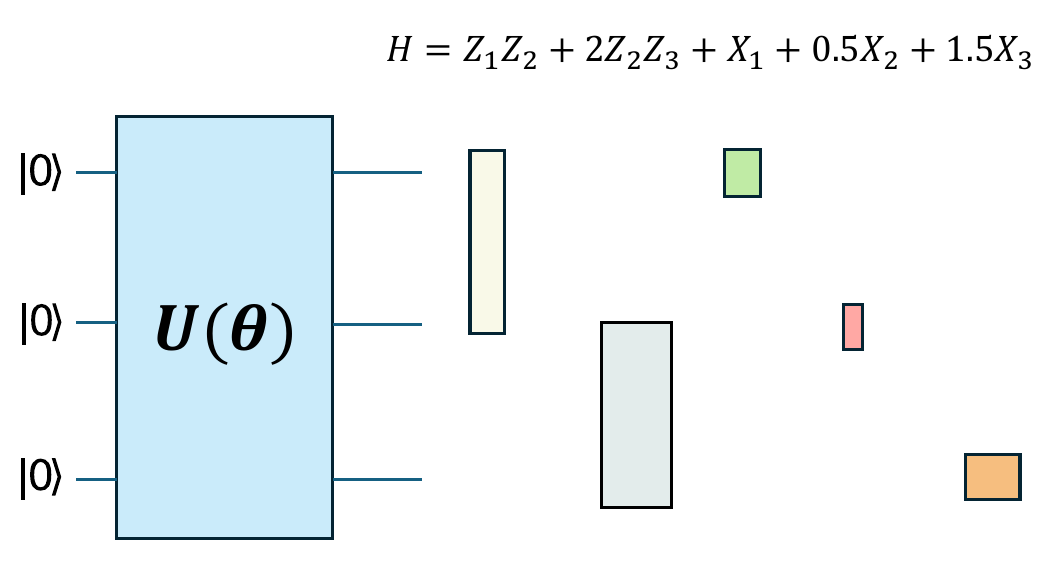}
  \caption{An illustration when the Hamiltonian is $H=Z_1Z_2+2Z_1Z_2+X_1+0.5X_2+1.5X_3$. The small rectangles indicate that the corresponding Hamiltonian term acts on that subsystem. And the width of rectangles reflects the magnitude of each coefficient $h_m$. Each subsystem contributes differently to the output due to the coefficients, suggesting that they should be assigned different weights in the optimization process.}
  \label{fig:subsystem}
\end{figure*}

Consider a $k$-local Hamiltonian $H=\sum_m h_mH_m$ and its expectation value $\tr(\rho_{\theta}H)=\sum_m h_m \tr(\rho_m H_m)$, where each $H_m$ is a Pauli string that acts non-trivially on a subsystem state $\rho_m$ of $k$ qubits. Intuitively, the contributions of each subsystem $\rho_m$ are different due to the associated weights $h_m$. An illustration (shown in \cref{fig:subsystem}) is the $3$-qubit toy Hamiltonian $H=Z_1Z_2+2Z_2Z_3+X_1+0.5X_2+1.5X_3$. The contributions of the five subsystems, corresponding to the five terms in the Hamiltonian, are weighted by their respective coefficients. Therefore, in the optimization process, a potential improvement is to also assign appropriate weights to each subsystem, rather than treating them equally as in standard QNG. In addition, as the total system size increases, the parameter sensitivity within each subsystem may differ significantly from that of the entire system. In such cases, the quantum Fisher information matrix $F$ of the entire system may struggle to capture the parameter sensitivity of each subsystem. To address all these aspects that are overlooked in QNG, we propose the WA-QNG method.

\subsection{Method Formulation} \label{MF}
In this subsection, we formalize the WA-QNG method. Suppose the target Hamiltonian to solve is $H=\sum_m h_m H_m$. Then the update formula of WA-QNG is then given by:

\begin{equation}
\theta^{(k+1)}=\theta^{(k)}-\eta W^{+} \nabla f(\theta^{(k)})
\label{eq:wa-qng}
\end{equation}

where $W=\frac{2}{\sum_m h_m^2}\sum_{m}h_m^2T_{m}$, and $T_m$ represents the Hilbert-Schmidt metric tensor \cite{koczor2022quantum1, koczor2022quantum2} of the $m$-th subsystem corresponding to the observable term $H_m$. The $i$-th row $j$-th column element of each $T_m$ is that $(T_m)_{ij}=\tr(\partial_i \rho_m\partial_j \rho_m)$. Because the coefficient $h_m$ can be negative, we use square-weighted summation instead of direct-weighted summation. The prefactor $\frac{2}{\sum_m h_m^2}$ is to make WA-QNG consistent with QNG: If each term $H_m$ in the Hamiltonian acts globally on the whole quantum system, then WA-QNG will reduce to QNG and $W=2T=F$. The proof of this equivalence is detailed in \cref{DG}. From this perspective, WA-QNG can be regarded as a more generalized form of the original QNG.

In the following, we further motivate and explain why the matrix $W$, namely the weighted average of the Hilbert-Schmidt metric tensors of each subsystem, is chosen in the updated formula of WA-QNG, and also why it is expected to perform well from three different interpretative perspectives. 

\subsection{Intuitive Interpretation} \label{WI}
The quantum Fisher information matrix $F$ of the entire system is used in the optimization step of QNG. Mathematically, because $F$ is unrelated to the index $m$, it can be rewritten as:
\begin{align}
    F &= 1\cdot F \notag \\
      &= \frac{1}{\sum_mh_m^2}\sum_mh_m^2F
\label{eq:QNG}
\end{align}
where $h_m$ is the $m$-th coefficient of the $k$-local Hamiltonian $H=\sum_m h_m H_m$. However, since the $m$-th observable term $H_m$ and coefficient $h_m$ are only associated with the subsystem $\rho_m$, an intuitive way to address the different contributions of each subsystem is to replace $F$ in \cref{eq:QNG} with $F_m$, the quantum Fisher information matrix of the corresponding subsystem $\rho_m$. Therefore, we define a new matrix $\hat{F}$:
\begin{equation}
    \hat{F} = \frac{1}{\sum_mh_m^2}\sum_mh_m^2F_m
    \label{eq:W-QNG}
\end{equation}
Compared to \cref{eq:QNG}, \cref{eq:W-QNG} incorporates a weighted sum of the quantum Fisher information matrices of individual subsystems. Here, $h_m^2$ serves as a weight to quantify the influence of each subsystem. By taking a weighted sum over the quantum Fisher information matrix of each subsystem $\rho_m$ corresponding to $H_m$, this formulation explicitly accounts for the different weights and contributions of each subsystem.

However, since each subsystem state will be a mixed state in general, and as noted in \cite{liu2020quantum, beckey2022variational, koczor2022quantum1}, the estimation of the quantum Fisher information for a mixed state is generally a more computationally demanding task than for a pure state. Consequently, estimating the matrix $F_m$ required in \cref{eq:W-QNG} is computationally demanding. To deal with this problem, references \cite{koczor2022quantum1, koczor2022quantum2} propose using the Hilbert-Schmidt metric tensor $T$ as an approximation of the quantum Fisher information $F$ in QNG for a mixed state, where $F \approx 2T$ when the mixed state is close to being pure and does not change dramatically with parameters. We also demonstrate this approximation relation in \cref{AQFI}. Therefore, we can approximate each $F_m$ in \cref{eq:W-QNG} using the Hilbert-Schmidt metric tensor $T_m$:
\begin{equation}
    W = \frac{2}{\sum_m h_m^2} \sum_m h_m^2 T_m
    \label{eq:WA-QNG}
\end{equation}
here we exactly obtain the matrix $W$ used in \cref{eq:wa-qng} of the update rule of WA-QNG. \cref{eq:W-QNG} introduces a weighted average to account for the relative importance of each subsystem. From \cref{eq:W-QNG} to \cref{eq:WA-QNG}, the Hilbert-Schmidt metric tensor is employed to approximate the quantum Fisher information matrix of each subsystem state. Hence, this is where the name of our method WA-QNG: \textit{Weighted Approximate Quantum Natural Gradient Descent} comes from.

\subsection{Optimization Interpretation} \label{GI}
The constrained optimization problem defined in  \cref{qng_problem} uses the fidelity distance $D_F$ as the distance metric. As discussed in \cref{qngd}, using $D_F$ instead of the Euclidean distance in the parameter space leads to the derivation of the QNG update formula. However, the distance metric $D_F$ for the entire quantum system still does not account for the different weights of each subsystem with respect to the observable terms. To capture this characteristic, we introduce the following distance:
\begin{equation}
    D_{W}\big(\rho(\theta+\delta), \rho(\theta)\big)= \frac{2}{\sum_m  h_m^2  }\sum_m  h_m^2  \Vert \rho_m(\theta+\delta)-\rho_m(\theta) \Vert_2^2
    \label{waqngdis}
\end{equation}
where $h_m$ is the $m$-th coefficient of the $k$-local Hamiltonian $H=\sum_m h_mH_m$, and $\rho_m$ represents the subsystem state of $\rho(\theta)$ corresponding to the $m$-th subsystem. Additionally, the coefficient $h_m^2$ serves as a weight in the weighted averaging process of the $2$-norm distance, thereby accounting for the different contributions of each subsystem. Hence, similar to \cref{qng_problem}, we define the following constrained optimization problem for each update step:

\begin{equation}\label{waqng_problem}
\begin{array}{cl}
\min\limits_{\delta} &\; f(\theta + \delta) \\
\mbox{s.t.} &\; D_{W}\big(\rho(\theta), \rho(\theta +\delta)\big)\leq \varepsilon
\end{array}
\end{equation}

From the constrained optimization problem defined above, we can derive the same update rule of WA-QNG as given in \cref{eq:wa-qng}. Thus, we establish WA-QNG from the optimization perspective. The detailed derivation from this optimization problem to WA-QNG is provided in \cref{GEO}.

\subsection{Geometric Interpretation} \label{PB}
The interpretation in the previous subsection can be further explained from the perspective of Riemannian geometry. In general, consider a function $F:\Theta \rightarrow \mathcal{M}$ from the parameter space $\Theta \subseteq \mathbb{R}^A$ to a Riemannian manifold $\mathcal{M}$ equipped with a Riemannian metric $g_{\mathcal{M}}$. A pullback metric \footnote{Strictly, only when $F$ is an immersion, the pullback metric defined in \cref{eq:pullback} is guaranteed actually to be a Riemannian metric. However, the pullback metric defined in \cref{eq:pullback} is always well-defined and ensures that the update formula in \cref{rieman} works in general.} $g$ on $\Theta$ is induced by function $F$, which is defined as \cite{lee2018introduction}:
\begin{align}
    g_{ij}&=(F^*g_{\mathcal{M}})(d \theta_i, d \theta_j) \notag \\
    &= g_{\mathcal{M}}(\partial_i F(\theta), \partial_j F(\theta))
    \label{eq:pullback}
\end{align}

Then, the Riemannian gradient descent \cite{amari1998natural, stokes2020quantum, gacon2021simultaneous} with pullback metric can be defined for $\mathcal{M}$:
\begin{equation}
    \theta^{(k+1)} = \theta^{(k)} - \eta g(\theta)^{+}\nabla f(\theta^{(k)})
    \label{rieman}
\end{equation}

Now, consider a quantum state $\rho(\theta)$ prepared by a parameterized circuit. For QNG, the quantum circuit defines a mapping function $F:\mathbb{R}^A \rightarrow \mathbb{H}^{N}$, where $\mathbb{H}^{N}$ represents the $N\times N$ Hermitian matrix space. In this view, QNG can be considered to work with the pullback metric of the Frobenius inner product $g_{\mathbb{H}}(\rho, \sigma)=\tr(\rho\sigma)$ defined on Hermitian matrix space:
\begin{align}
    g_{ij}(\theta)&= (F^*g_{\mathbb{H}})(d \theta_i, d \theta_j)  \notag \\
    &= g_{\mathbb{H}}(\partial_i \rho(\theta), \partial_j\rho(\theta)) \notag \\
    &= \tr(\partial_i \rho(\theta)\partial_j\rho(\theta))
\end{align}

which is consistent with the update rule of QNG in~\cref{eq:qng}, up to a constant factor $2$. In WA-QNG, to address the different weights and contributions of each subsystem, we view the mapping function $F$ as:
\begin{equation}
    \theta \xrightarrow{\text{F}} q(\theta)=\frac{1}{\sqrt{\sum_mh_m^2}}[h_1\rho_1(\theta),\cdots ,h_M\rho_M(\theta)]
\end{equation}

where the $q(\theta)$ is a point in the product space $\mathbb{H}_1\times\cdots\times\mathbb{H}_M$ where each subsystem $\rho_m \in \mathbb{H}_m$. This product space is equipped with an inner product $\langle (\rho_1, \ldots, \rho_M), (\sigma_1, \ldots, \sigma_M) \rangle = \sum_{m=1}^M \tr(\rho_m\sigma_m)$. Similarly, the pullback metric can also be obtained as:

\begin{align}
    g_{ij}(\theta) &= \langle \partial_i q(\theta), \partial_j q(\theta)\rangle \notag \\
    &=  \frac{1}{\sum_mh_m^2}\sum_m h_m^2 \tr(\partial_i\rho_m(\theta)\partial_j\rho_m(\theta)) 
\end{align}

which is consistent with the update rule of WA-QNG in \cref{eq:wa-qng}, also up to a constant factor $2$. Compared to QNG, since each subsystem is explicitly represented with its corresponding coefficient in the direct product space, the pullback Riemannian metric tensor is more likely to account for the different weights of each subsystem.

\section{Relation with Gauss-Newton Method} \label{OI}
As illustrated in \cite{martens2020new}, second-order optimization methods that leverage Fisher information can be interpreted as a generalized Gauss-Newton method. Here, we also demonstrate that the objective function \cref{eq:vqe} can be approximately transferred into a weighted non-linear least squares problem when each subsystem is close to being pure and does not change significantly with respect to parameters. Under this condition, we prove that WA-QNG is equivalent to the Gauss-Newton method for this non-linear least squares problem.

Let $\tilde{H_m}=-H_m$ and $\hat{H}_m=\frac{\tilde{H}_m}{h_m}$ for simplicity in the derivation. Also note that all constant factors, such as $\frac{1}{\sum_m h_m^2}$, do not affect the optimization formulation, as they can ultimately be absorbed into the learning rate. For simplicity, we use the symbol $\Leftrightarrow$ to represent two minimization problem are equivalent up to a constant factor. Starting from the optimization problem in \cref{eq:vqe}, we can perform the following transformation:
\begin{align}
    & \min_{\theta}  \tr(\rho(\theta)H) \notag \\
    &\Leftrightarrow \min_{\theta} \frac{1}{\sum_m h_m^2}\sum_m\tr(h_m\rho_m(\theta)H_m) \notag \\
    &\Leftrightarrow \min_{\theta} \frac{1}{\sum_m h_m^2} \sum_m -\tr(h_m\rho_m(\theta)\tilde{H}_m) \notag \\
    &\approx \min_{\theta} \frac{\sum_m \big(\tr(h_m^2\rho_m^2(\theta))-2\tr(h_m\rho_m(\theta)\tilde{H}_m) + \tr(\tilde{H}_m^2)\big) }{\sum_m h_m^2} 
    \label{mns1}
\end{align}

Note that the third transformation is an approximate one, where two additional terms, $\tr(h_m^2\rho_m^2)$ and $\tr(\tilde{H}_m^2)$, are added into the summation. The latter is a constant so it does not affect the optimization. For the former, we leverage the assumption that each subsystem is close to being pure and does not change significantly with respect to the parameters. Under this condition, the term $\tr(h_m^2\rho_m^2)$ remains close to a constant $h_m^2$, while the term $2\tr(h_m\rho_m\tilde{H}_m)$ dominants the gradient in the optimization problem. When $\rho_m$ is exactly pure, the approximate transformation becomes exact. We then continue the transformation:

\begin{align}
    &\min_{\theta} \frac{\sum_m \big(\tr(h_m^2\rho_m^2(\theta))-2\tr(h_m\rho_m(\theta)\tilde{H}_m) + \tr(\tilde{H}_m^2)\big)}{\sum_m h_m^2}  \notag \\
    &\Leftrightarrow \min_{\theta} \frac{1}{\sum_m h_m^2} \sum_m  \Vert h_m\rho_m(\theta) - \tilde{H}_m \Vert^2_2 \notag \\
    &\Leftrightarrow \min_{\theta} \frac{1}{\sum_m h_m^2} \sum_m h_m^2 \Vert \operatorname{vec}(\rho_m(\theta)) -\operatorname{vec}(\hat{H}_m)  \Vert^2
    \label{mns2}    
\end{align}

In the final expression of \cref{mns2}, we observe that the original problem is transformed into a non-linear least squares problem. The update formula of Gauss-Newton method \cite{nocedal1999numerical} for such a non-linear least squares problem with weights is given by:
\begin{equation}
    \theta^{(k+1)}=\theta^{(k)}-\eta (J_r^TJ_r)^{-1} J_r^T \vec{r}(\theta^{(k)})
    \label{gnm}
\end{equation}
$\vec{r}(\theta)$ is often referred to as the residual vector which is defined in \cref{resi} in our case, and $J_r$ is the Jacobian of the residual with respect to the parameter $\theta$.
\begin{align}
    &\vec{r}(\theta) =  
    \frac{[h_1(\operatorname{vec}(\rho_1)-\operatorname{vec}(\hat{H}_1)) ,\cdots,  h_M( \operatorname{vec}(\rho_M)-\operatorname{vec}(\hat{H}_M))]^T}{\sqrt{\sum_mh_m^2}}
    \label{resi}    
\end{align}

 Note that, as the objective function is defined by \cref{mns2}, where $ f(\theta)=\frac{1}{\sum_m h_m^2}\sum_m h_m^2 \Vert \operatorname{vec}(\rho_m) -\operatorname{vec}(\hat{H}_m)  \Vert^2=\vec{r}^T\vec{r}$, \cref{gnm} can be rewritten into:
\begin{equation}
    \theta^{(k+1)}=\theta^{(k)}-\eta (2J_r^TJ_r)^{-1} \nabla f(\theta^{(k)})
    \label{gnm2}
\end{equation}
One can verify that the matrix $W$ defined in WA-QNG satisfies the relation: $W=2J_r^{T}J_r$. Hence, the update rule in \cref{gnm2} is fully equivalent to the update rule of WA-QNG in \cref{eq:wa-qng}. The details of the relationship between the matrix $W$ and the Gauss-Newton method are provided in \cref{RGCI}.

Thus, we have demonstrated that WA-QNG is approximately equivalent to the Gauss-Newton method for a nonlinear least squares problem, under the assumption that each subsystem is close to being pure and does not vary significantly with respect to the parameters. Under this condition, WA-QNG is expected to inherit the favorable properties of the Gauss-Newton method and has the potential to outperform ordinary gradient descent.

\section{Complexity Analysis} \label{CAS}
In this section, we analyze the computational complexity in each optimization step of WA-QNG in comparison with standard QNG.

For both methods, it is necessary to estimate the gradient and the metric tensor at each optimization step. The computational costs for the gradient $\nabla f$ are identical for both methods, so the main comparison lies in the computational cost of obtaining the metric $F$ in \Cref{eq:qng} and the weighted metric $W$ in \Cref{eq:wa-qng}. For simplicity, the computational complexity discussed in this section refers particularly to that of computing the metric tensor, as the complexity of computing the gradient is the same for both methods.

For standard QNG, it is necessary to estimate the elements of the matrix $F$, so the number of elements to be estimated scales with $v^2$, where $v$ is the number of parameters in the circuit. For WA-QNG, each metric tensor of the subsystem in the weighted summation must be estimated, so the total number of elements to be estimated scales with $mv^2$, where $m$ is the number of local terms in the Hamiltonian. For both methods, each element in the corresponding matrices can be estimated using the Hadamard test or the Swap test \cite{koczor2022quantum1, anuar2024operator}, both of which require additional ancillary circuits. In this case, WA-QNG requires a larger number of tests in each optimization step compared to standard QNG.

In scenarios with limited circuit scale, where additional ancillary circuits are not feasible, the classical shadow technique \cite{huang2020predicting} can be employed to estimate the required elements of the matrices for both methods. We show that the complexity of obtaining each element in the metric tensor is $\mathcal{O}(2^k)$, where $k$ is the locality of the corresponding subsystem. The detailed derivation is provided in \cref{CS}. For standard QNG, the quantum Fisher information matrix corresponds to the entire system, involving all qubits, i.e., $k = n$. In contrast, for WA-QNG, each metric tensor $T_m$ in the weighted summation is associated only with a $k$-local subsystem. Therefore, the time complexity for estimating $F$ in QNG is $\mathcal{O}(v^2 \cdot 2^n)$, while that for estimating $W$ in WA-QNG is $\mathcal{O}(mv^2 \cdot 2^k)$.

In typical applications, the number of Hamiltonian terms $m$ does not grow exponentially with the number of qubits $n$. For example, in molecular systems, $m = \mathcal{O}(n^4)$ \cite{koczor2022quantum1}. Therefore, in scenarios with limited circuit scale, where the classical shadow technique is used to estimate metric tensors, WA-QNG is more shot-efficient than standard QNG in each optimization step. This is because the cost of WA-QNG scales exponentially with the locality $k$, while that of standard QNG scales exponentially with the total number of qubits $n$.

However, in both scenarios, each optimization step of WA-QNG remains more effective than that of standard QNG in terms of optimization performance. We will show this numerically in the next section.

\section{Numerical Results} \label{exp}
In this section, we present the results of the numerical simulations. First, we compare the overall performance of standard QNG and WA-QNG for the 1D Ising model and Heisenberg model in \cref{PC}. Additionally, to better evaluate whether the design of WA-QNG to capture the different weights of subsystems effectively improves upon standard QNG, we design numerics to examine the effects of subsystem weights and the number of qubits in \cref{WS} and \cref{NQ} respectively.
\subsection{Performance Comparison} \label{PC}
To evaluate the performance of WA-QNG, we test it alongside standard QNG on the variational quantum eigensolver for a 1D Ising model and Heisenberg model. Their Hamiltonians are given as follows, respectively:
\begin{equation}
    H=\sum_{\langle ij \rangle}Z_iZ_j + \sum_i X_i
    \label{Ising}
\end{equation}

\begin{equation}
    H=\sum_{\langle ij \rangle} (X_iX_j+Y_iY_j +Z_iZ_j)
\end{equation}

\begin{figure}[ht]
  \centering
  \includegraphics[width=0.45\textwidth]{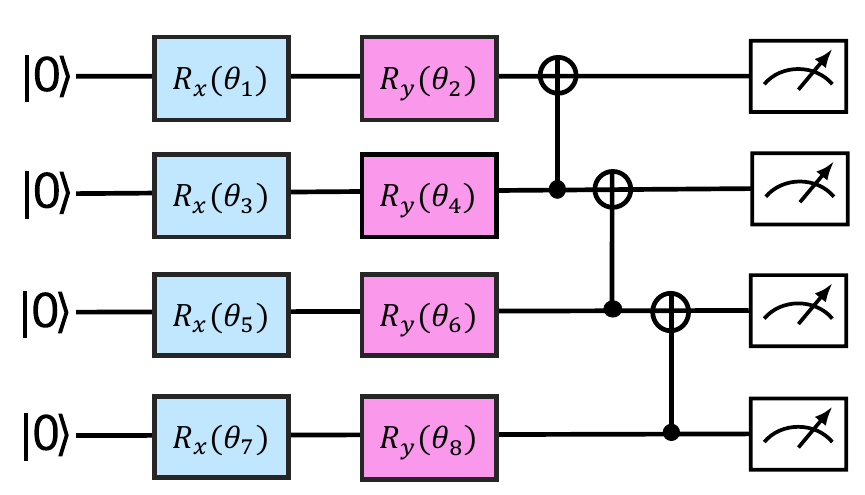}
  \caption{An example of a $4$-qubit EfficientSU2 circuit. It consists of single-qubit rotation gates $R_x$ and $R_y$, followed by a series of CNOT gates to enhance entanglement.}
  \label{fig:efficientsu}
\end{figure}

$\langle ij \rangle$ in the Hamiltonian denotes that the $i$-th and $j$-th qubits are the nearest neighbors. The variational quantum circuit used in our numerics is the widely used EfficientSU2, illustrated in \cref{fig:efficientsu} for the $4$-qubit, single-layer example. In our numerics, we evaluate QNG and WA-QNG for these two Hamiltonians on $10$-, $12$- and $14$-qubit cases for a single EfficientSU2 layer circuit. In our numerics, we evaluate QNG and WA-QNG for these two Hamiltonians on 10-, 12-, and 14-qubit cases with a single-layer circuit. In addition, for the 10-qubit case, we also evaluate the methods with multiple EfficientSU2 layers to examine the influence of the number of layers.

In our numerics, standard QNG and vanilla gradient descent are used as baseline methods for comparison with WA-QNG. Both WA-QNG and QNG require additional shots to estimate the metric tensor, whereas vanilla gradient descent does not. For a detailed comparison between standard QNG and vanilla gradient descent, please refer to \cite{stokes2020quantum}. As discussed in \cref{OI}, WA-QNG has a solid theoretical connection to the Gauss–Newton method when each subsystem is nearly pure. To achieve this, we propose initializing the variational circuit with small angles to ensure low entanglement at the start. Moreover, small-angle initialization is believed to help mitigate issues such as the Barren Plateau problem \cite{larocca2024review, zhang2022escaping}. In our numerics, each parameter is uniformly randomly selected from the intervals $[-1,1]$, which limits the angle magnitude to be relatively small while maintaining adequate randomness to evaluate the methods. To ensure a fair comparison, the learning rate for all methods is set to $0.02$, and the parameters are initialized identically across all methods. Each configuration is independently run 50 times, and the learning curves presented in the following sections are averaged over these runs with standard deviation shading. All numerics are conducted on classical simulators, where both expectation values and metric tensors can be tracked precisely.

\begin{figure*}[ht]
    \centering
    \begin{minipage}{0.325\textwidth}
        \centering
        \includegraphics[width=\linewidth]{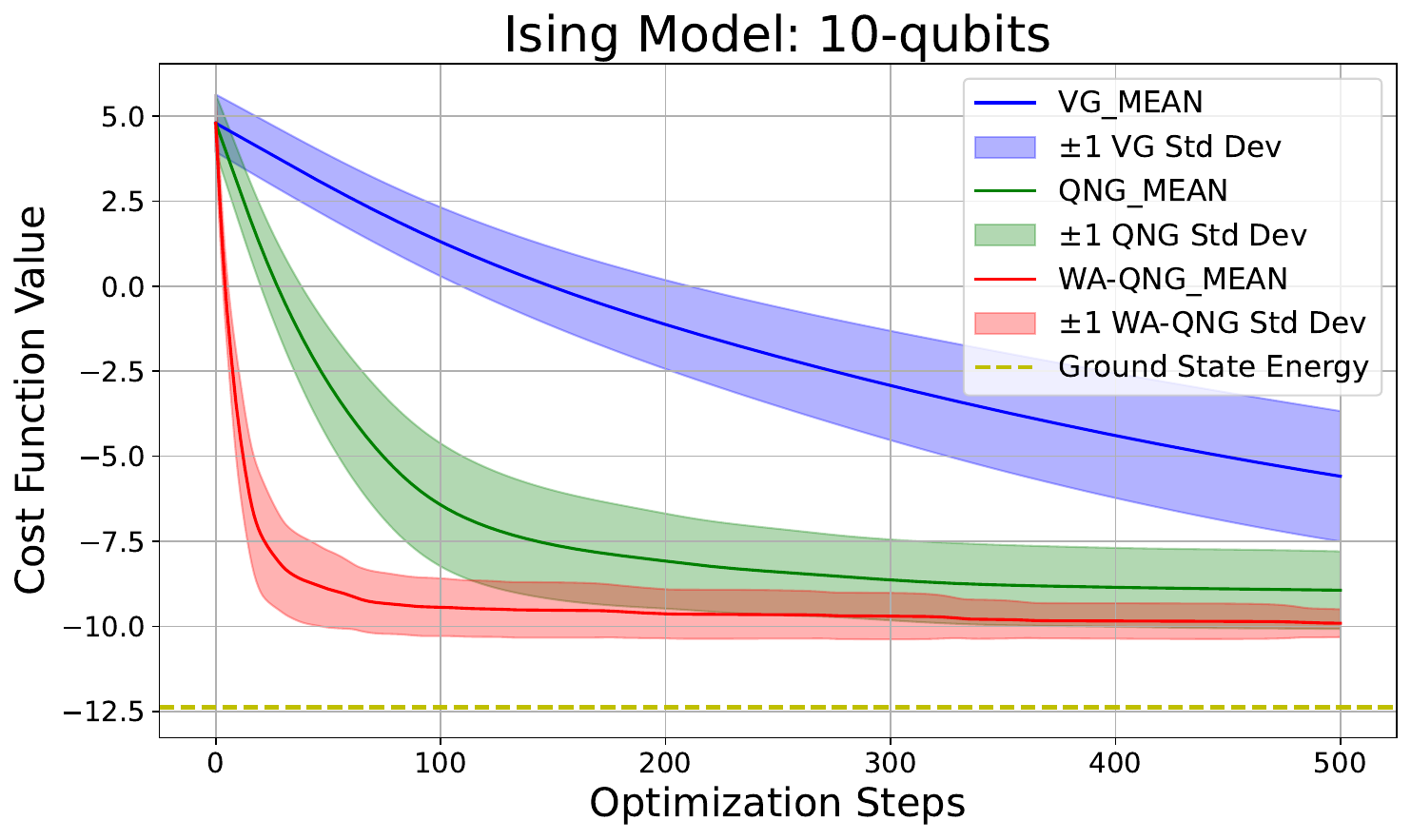}
    \end{minipage}
    \begin{minipage}{0.325\textwidth}
        \centering
        \includegraphics[width=\linewidth]{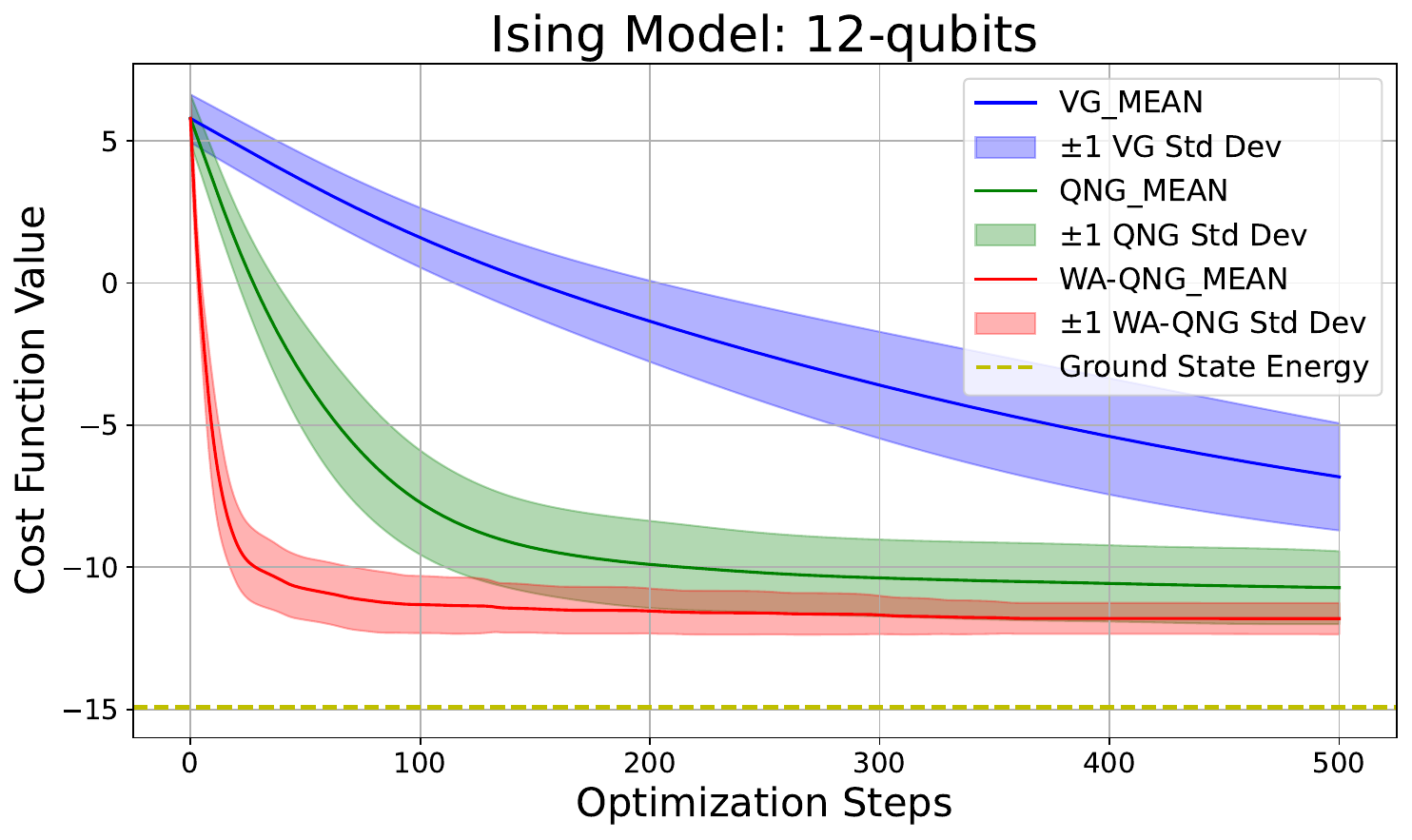}
    \end{minipage}
    \begin{minipage}{0.325\textwidth}
        \centering
        \includegraphics[width=\linewidth]{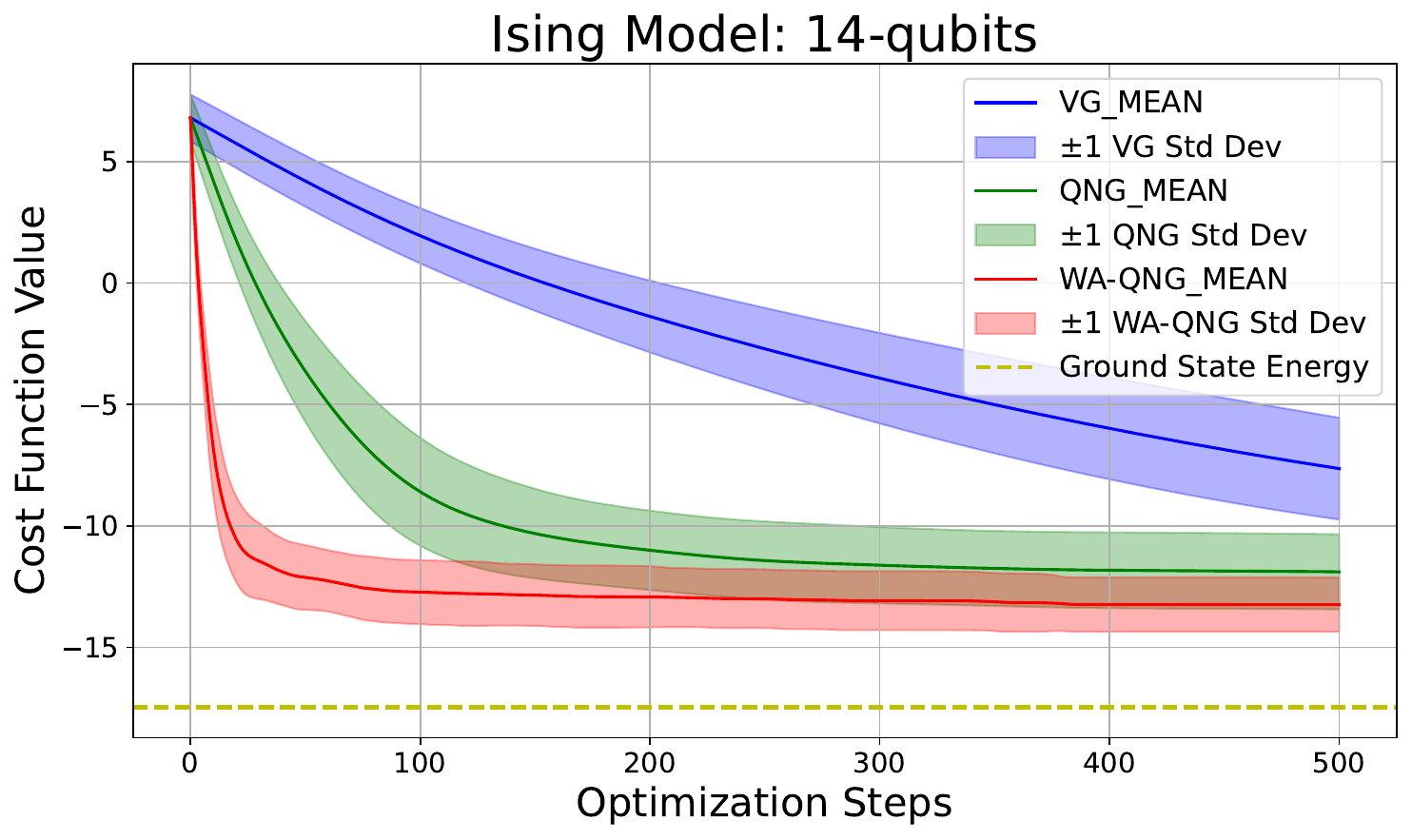}
    \end{minipage}
    \caption{The learning curves of the three methods for Ising model of $10$, $12$, $14$ qubits on 1-layer EfficientSU2 circuit.}
    \label{fig:learncurve_ising}
\end{figure*}

\begin{figure*}[ht]
    \centering
    \begin{minipage}{0.325\textwidth}
        \centering
        \includegraphics[width=\linewidth]{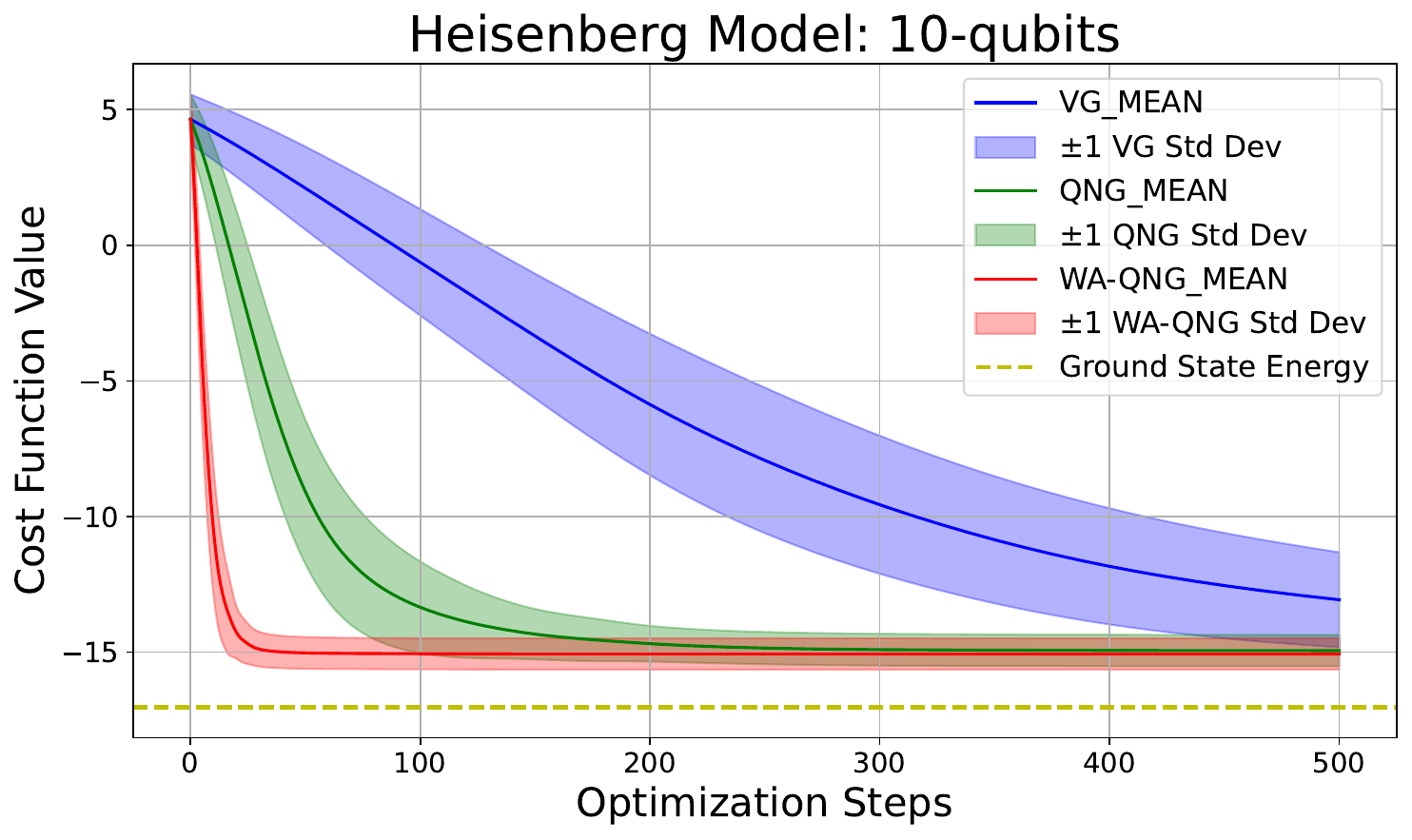}
    \end{minipage}
    \begin{minipage}{0.325\textwidth}
        \centering
        \includegraphics[width=\linewidth]{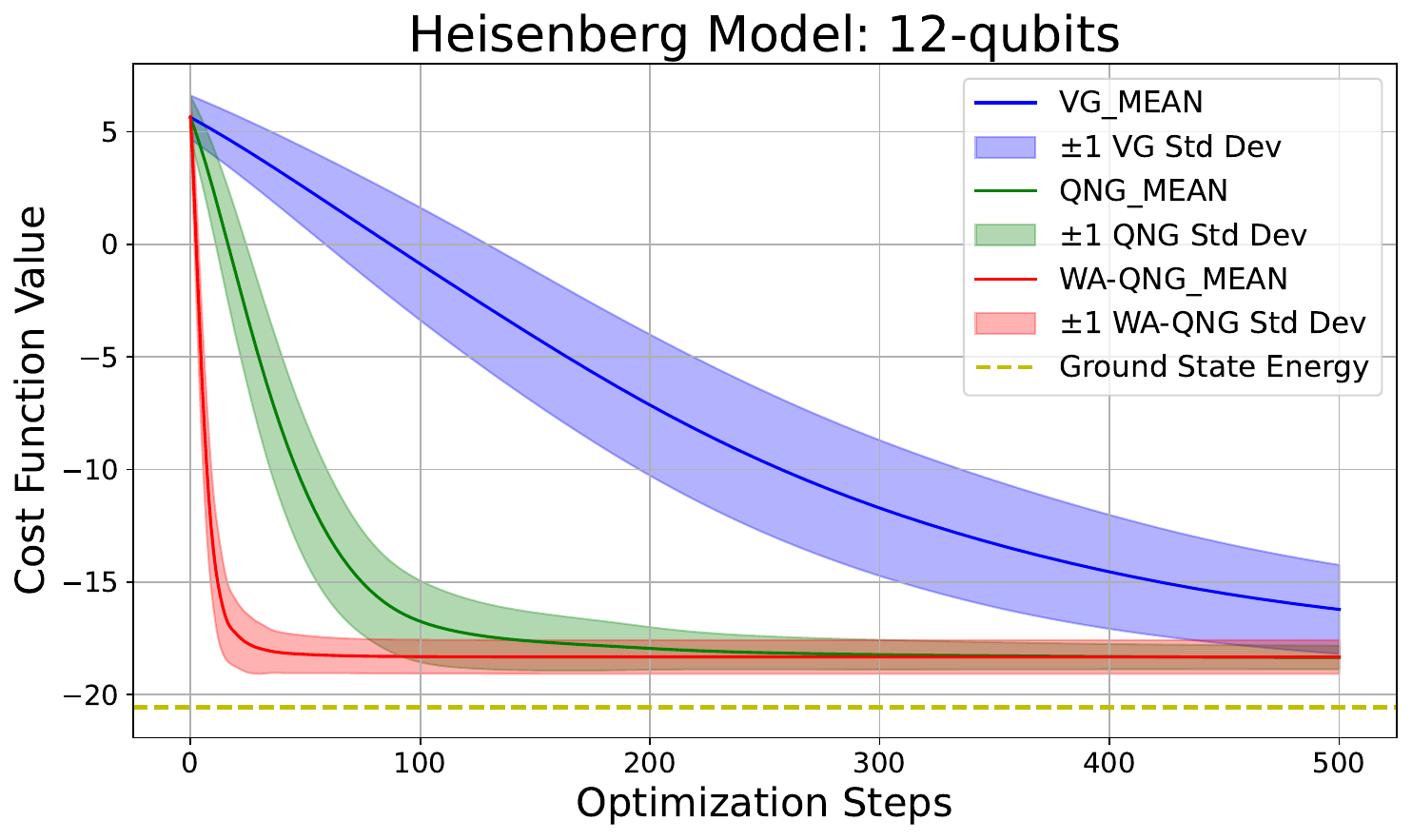}
    \end{minipage}
    \begin{minipage}{0.325\textwidth}
        \centering
        \includegraphics[width=\linewidth]{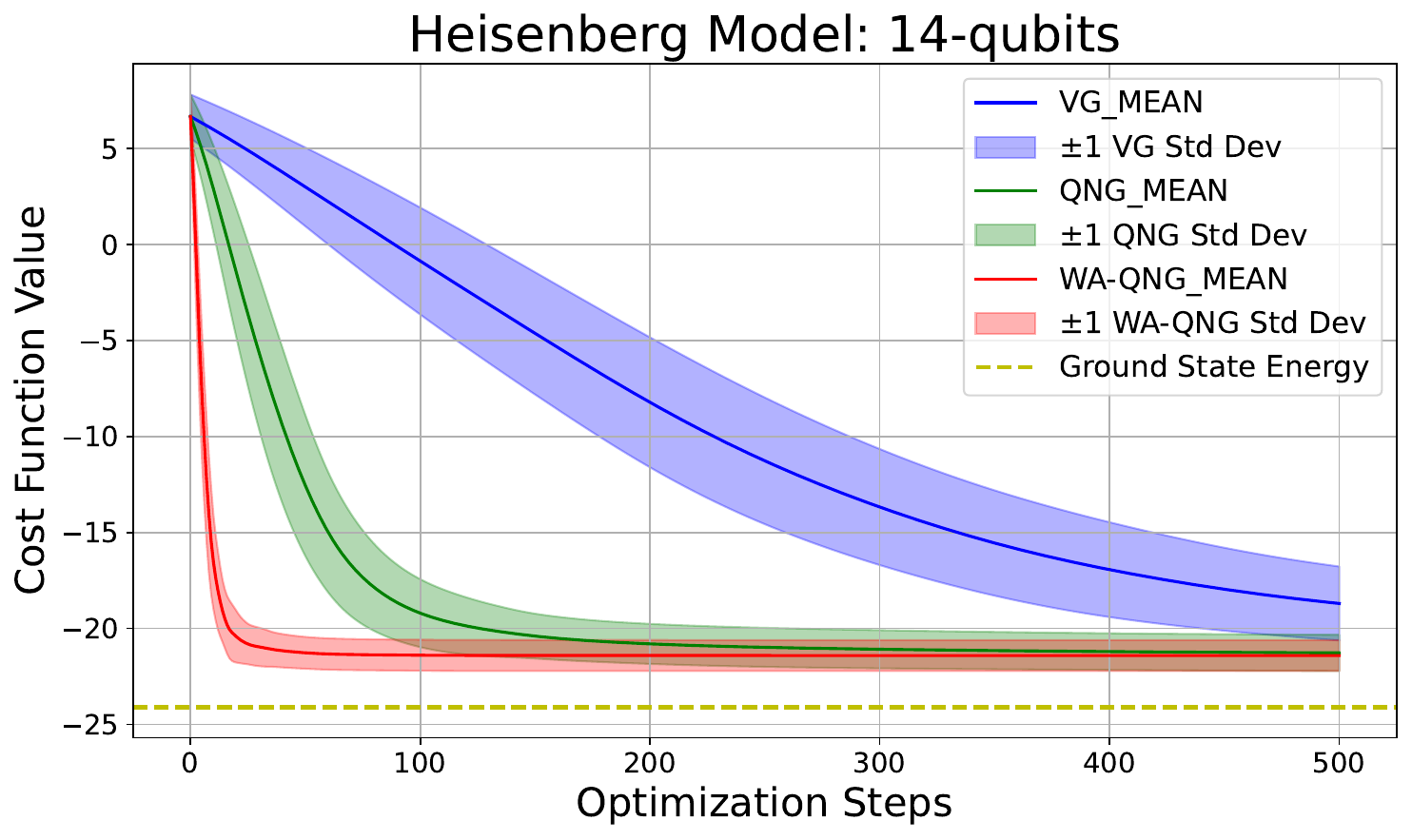}
    \end{minipage}
    \caption{The learning curves of the three methods for Heisenberg model of $10$, $12$, $14$ qubits on 1-layer EfficientSU2 circuit.}
    \label{fig:learncurve_hesen}
\end{figure*}

The learning curves of three methods for the 10-, 12-, and 14-qubit Ising and Heisenberg models with a single-layer circuit are shown in \cref{fig:learncurve_ising} and \cref{fig:learncurve_hesen}, respectively. Across all instances, WA-QNG exhibits a markedly faster convergence than both QNG and vanilla gradient descent, with the latter being the slowest and failing to converge within 500 optimization steps. These findings suggest that WA-QNG possesses a potential advantage in convergence speed compared to the other two methods. In the Ising model cases, WA-QNG attains an explicitly better average final convergence value than QNG, while in the Heisenberg model cases, the two methods yield almost the same final average convergence values, with WA-QNG still maintaining a slight advantage. This indicates that WA-QNG can provide superior, or at least competitive, performance compared to standard QNG in terms of the ability to escape local minima.

One may note that all three methods fail to converge to the true ground-state energy. This can be attributed to two reasons. First, the EfficientSU2 circuit used in the numerics is not guaranteed to include the ground state within its expressible space. Second, both WA-QNG and standard QNG may become trapped in local minima during the optimization process. Hence, the numerics in this work mainly aim to compare the relative performance between WA-QNG and standard QNG as gradient-based local search optimization methods. The absolute performance, related to global convergence to the true minimum, remains a great challenge in the optimization field.

\begin{figure*}[ht]
    \centering
    \begin{minipage}{0.325\textwidth}
        \centering
        \includegraphics[width=\linewidth]{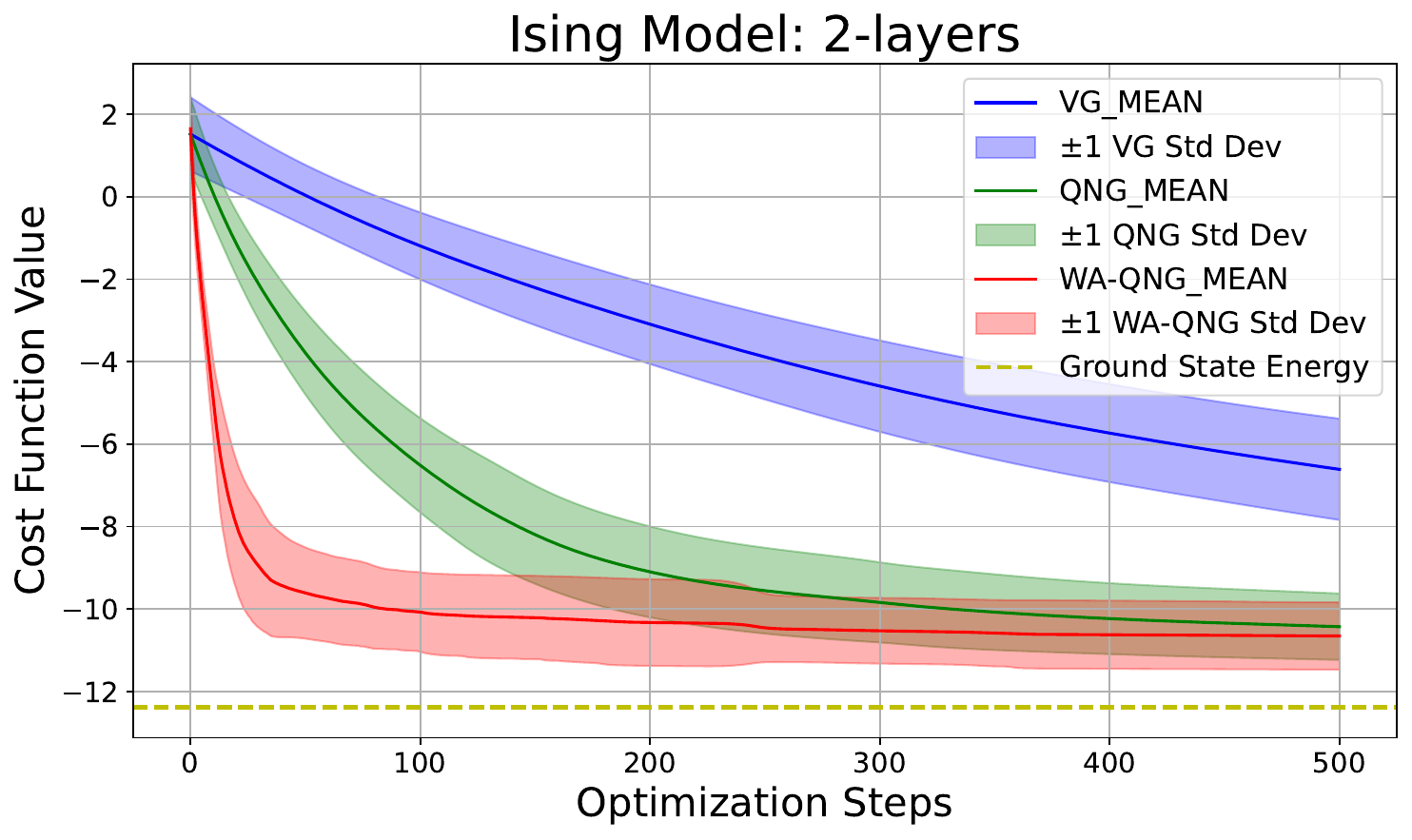}
    \end{minipage}
    \begin{minipage}{0.325\textwidth}
        \centering
        \includegraphics[width=\linewidth]{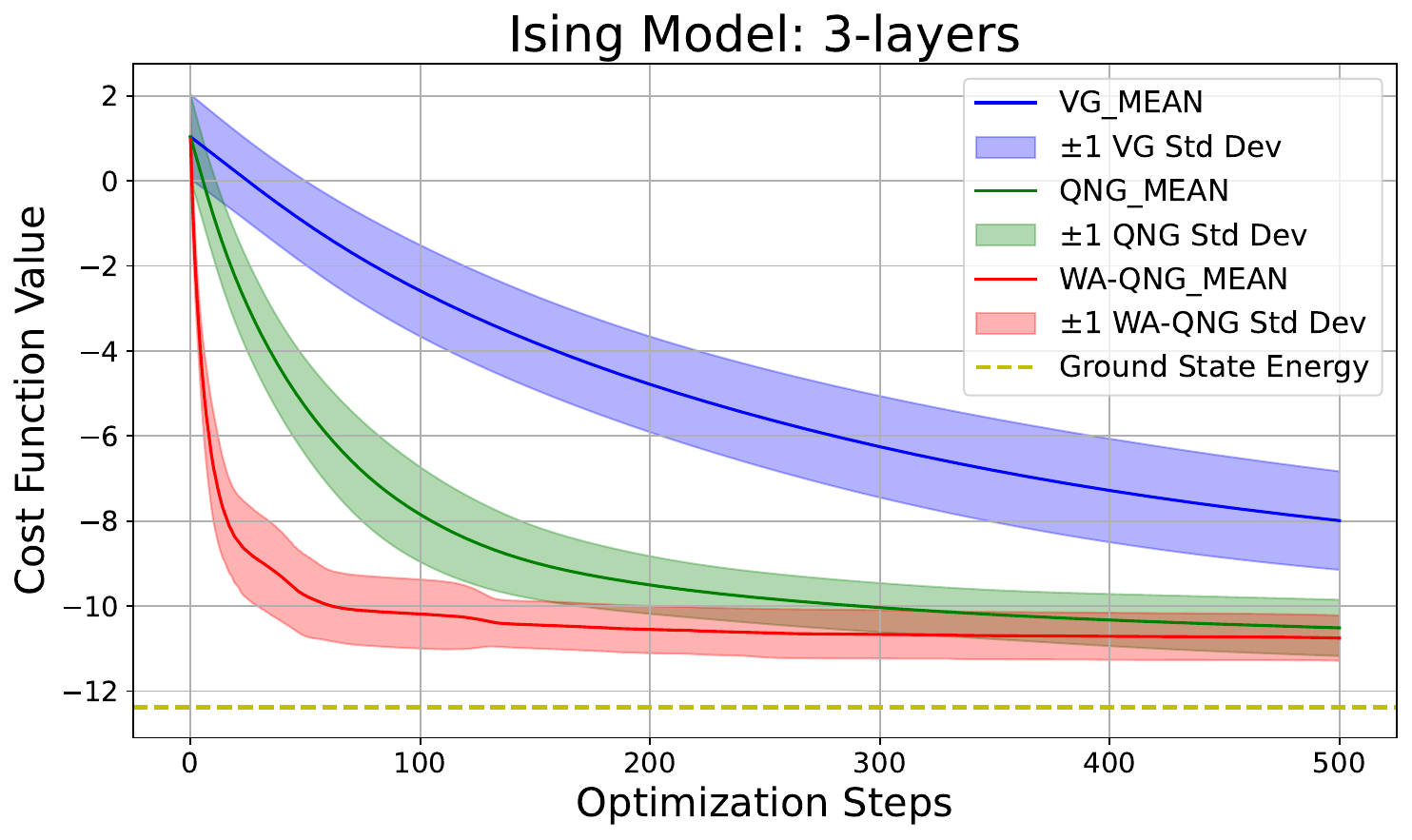}
    \end{minipage}
    \caption{The learning curves of the three methods for Ising model of $10$ qubits on $2$, $3$-layer EfficientSU2 circuit.}
    \label{fig:learncurve_im}
\end{figure*}

\begin{figure*}[ht]
    \centering
    \begin{minipage}{0.325\textwidth}
        \centering
        \includegraphics[width=\linewidth]{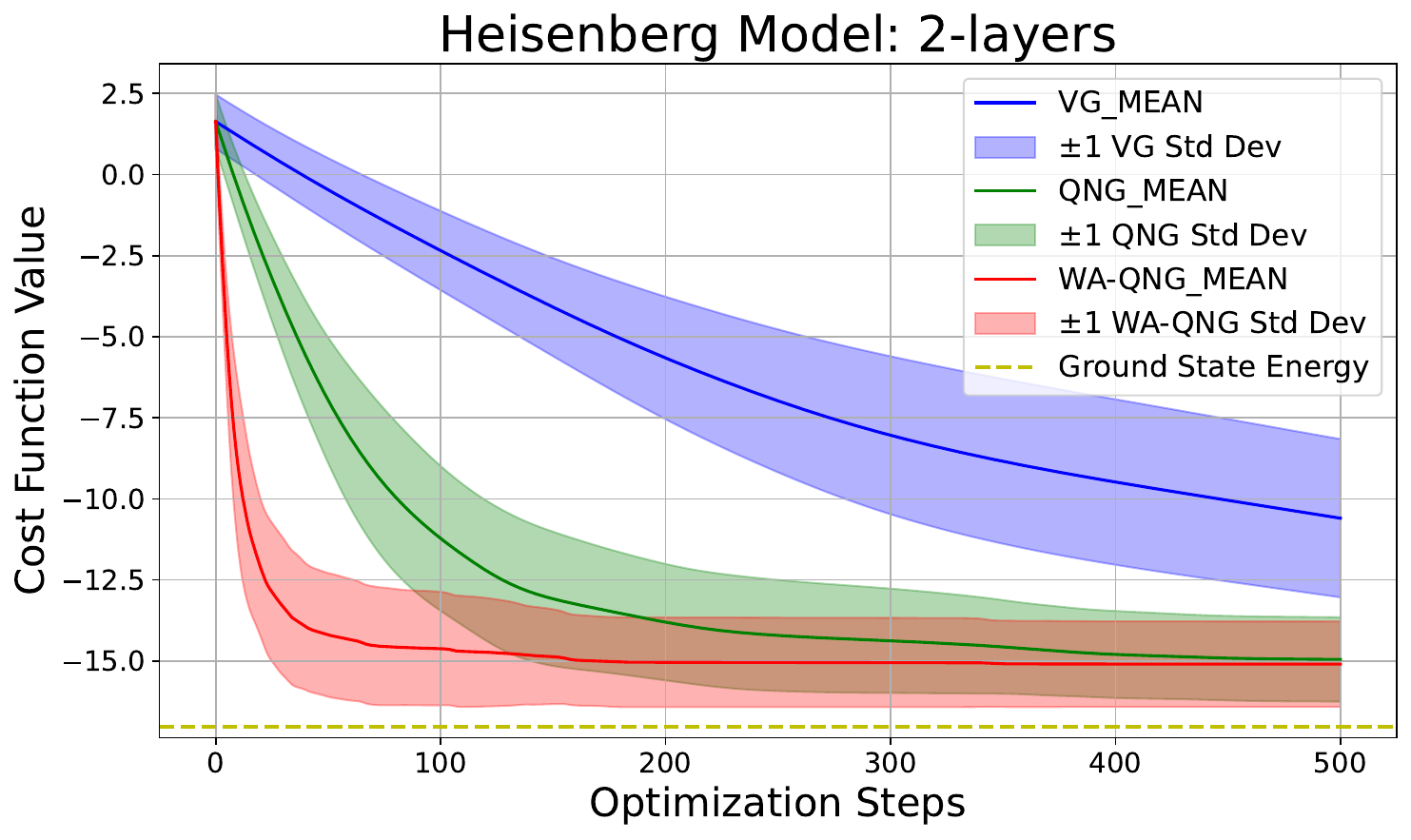}
    \end{minipage}
    \begin{minipage}{0.325\textwidth}
        \centering
        \includegraphics[width=\linewidth]{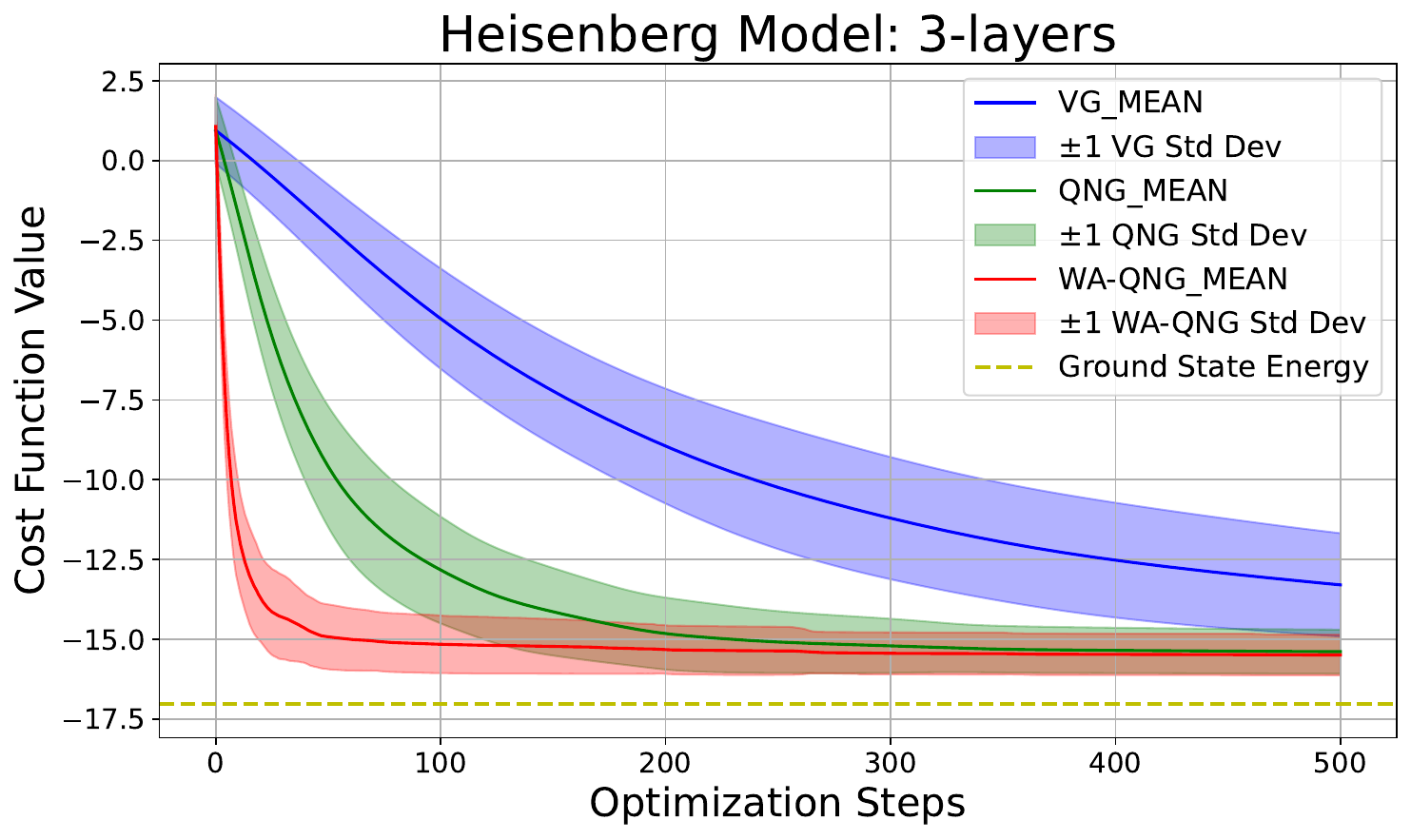}
    \end{minipage}
    \caption{The learning curves of the three methods for Heisenberg model of $10$ qubits on $2$, $3$-layer EfficientSU2 circuit.}
    \label{fig:learncurve_hm}
\end{figure*}

The learning curves of the three methods for the 10-qubit Ising and Heisenberg models with multi-layer circuits are shown in \cref{fig:learncurve_im} and \cref{fig:learncurve_hm}, respectively. The results indicate that the potential advantages discussed in the single-layer case still hold when the number of layers increases. Furthermore, both WA-QNG and standard QNG reach a sightly better average convergence value compared to the single-layer circuit-- although increasing the number of layers makes the optimization landscape more complex, it also enlarges the expressive space of the variational circuit, which may contain better solutions with lower cost function values. This result also suggests that WA-QNG can still perform well in a more complicated landscape.

\begin{figure*}[ht]
    \centering
    \begin{minipage}{0.325\textwidth}
        \centering
        \includegraphics[width=\linewidth]{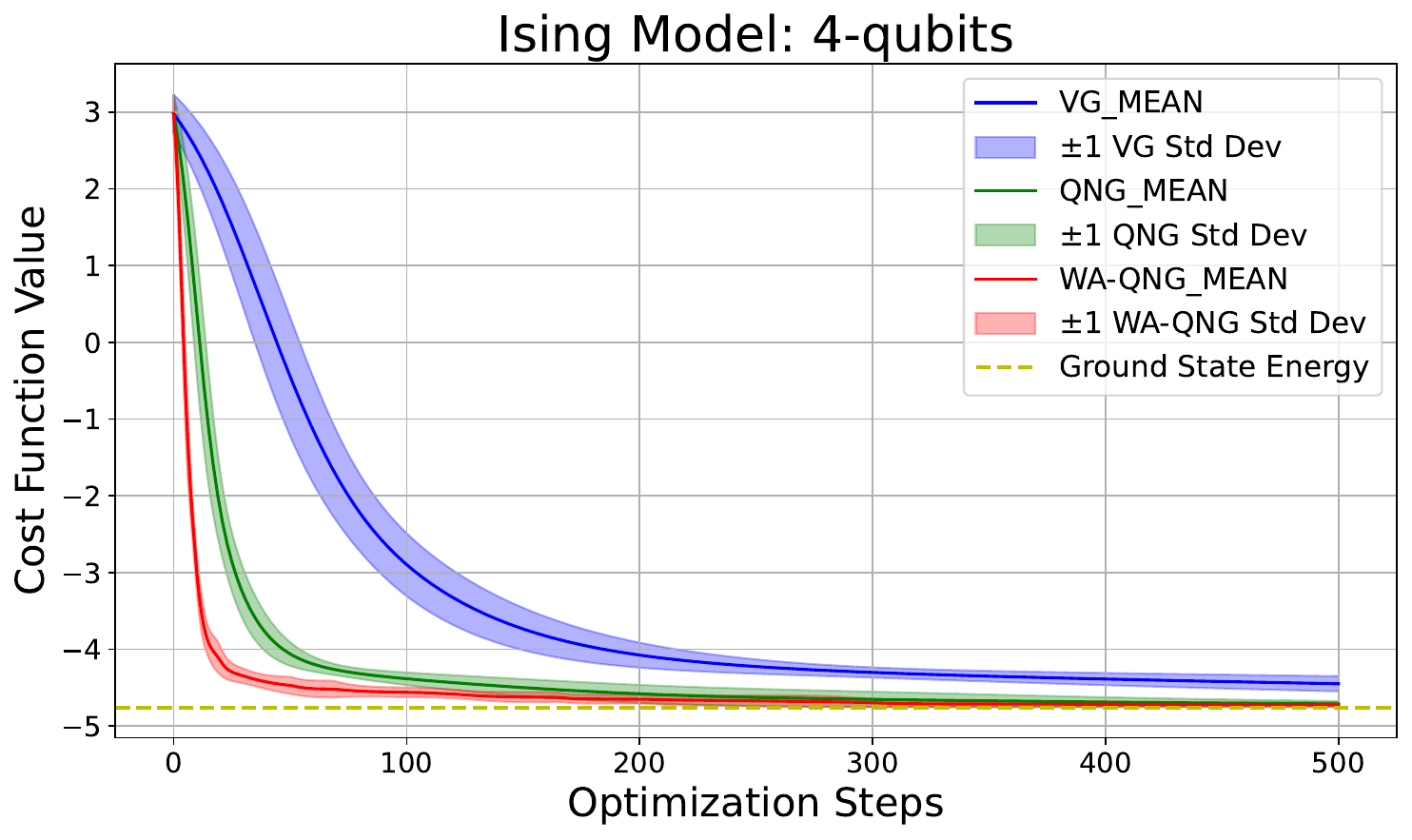}
    \end{minipage}
    \begin{minipage}{0.325\textwidth}
        \centering
        \includegraphics[width=\linewidth]{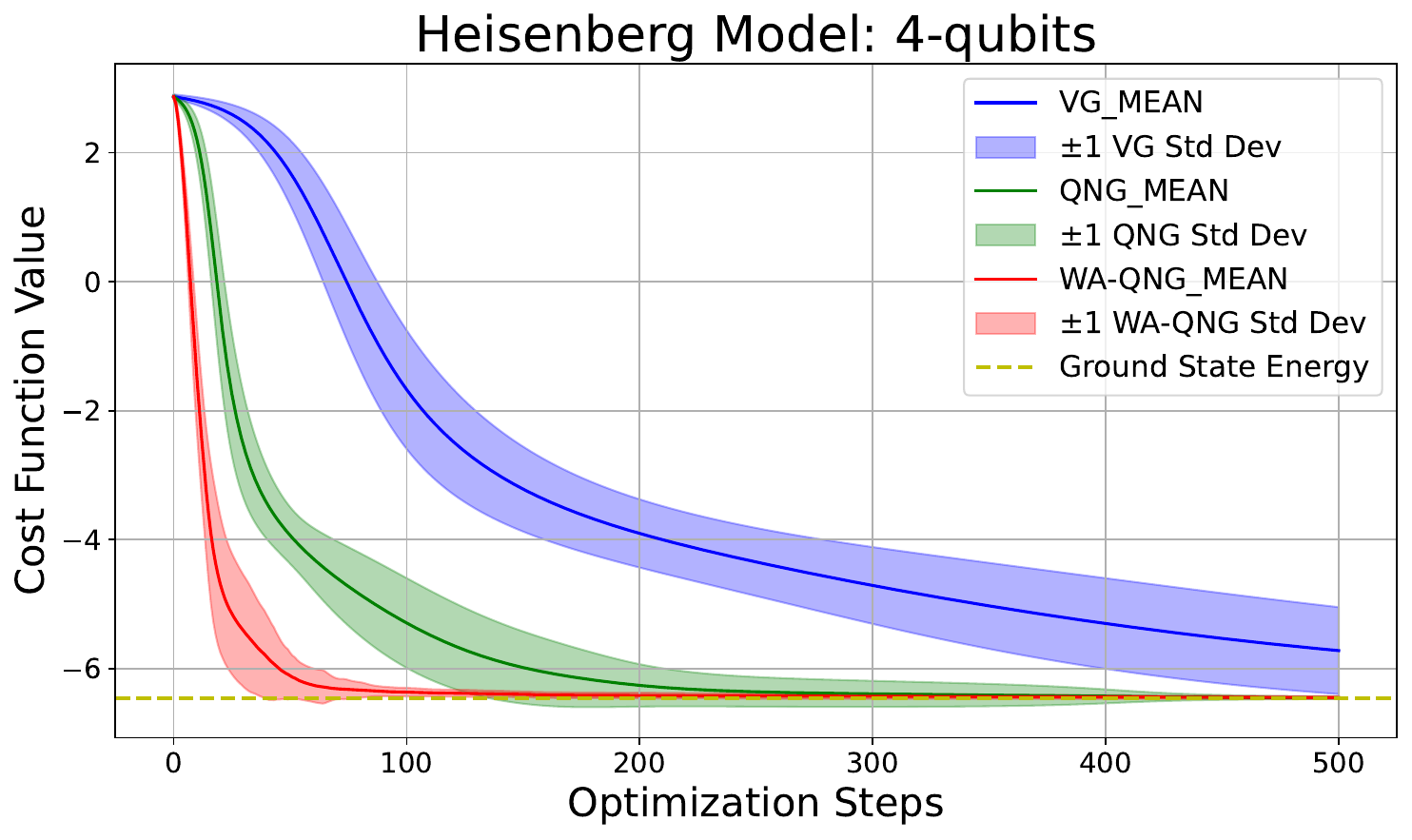}
    \end{minipage}
    \caption{The learning curves of the three methods for Ising and Heisenberg model of $4$ qubits on $4$-layer EfficientSU2 circuit.}
    \label{fig:4layer}
\end{figure*}

In addition, we also provide numerical simulations for the four-qubit Ising and Heisenberg models using a four-layer EfficientSU2 circuit, for which the ground state lies within the search space, as a supplement to the numerics discussed above. As shown in \Cref{fig:4layer}, both WA-QNG and QNG converge to the ground-state energy in this setting, where the circuit is sufficiently expressive, and WA-QNG still exhibits a faster convergence rate. In contrast, vanilla gradient descent remains trapped in a local minimum and fails to reach the ground state. This result further highlights the importance of the choice of optimizer in VQE: even when the ansatz is sufficiently expressive, a suboptimal optimizer may lead the optimization process to local minima and thus limit overall performance.

\subsection{Weights of Subsystems} \label{WS}
To better understand how accounting for the different weights of each subsystem in WA-QNG plays a central role in improving optimization performance compared to standard QNG, we conduct numerics using the following 3-qubit toy Ising model, with $\alpha$ set to $0.8$, $0.6$, $0.4$ and $0.2$:
\begin{equation}
   H=Z_1Z_2+Z_2Z_3+\alpha X_1+(3-2\alpha)X_2+ \alpha X_3
   \label{weights}
\end{equation}

As $\alpha$ decreases from $0.8$ to $0.2$,  the weights of the subsystems become increasingly unbalanced, with the contribution of the subsystem associated with the second qubit to the output growing larger. Therefore, if incorporating subsystem weights into the optimization is really effective, WA-QNG is expected to exhibit increasingly better performance compared to QNG as $\alpha$ decreases.

\begin{figure*}[ht]
  \centering
  \includegraphics[width=0.45\textwidth]{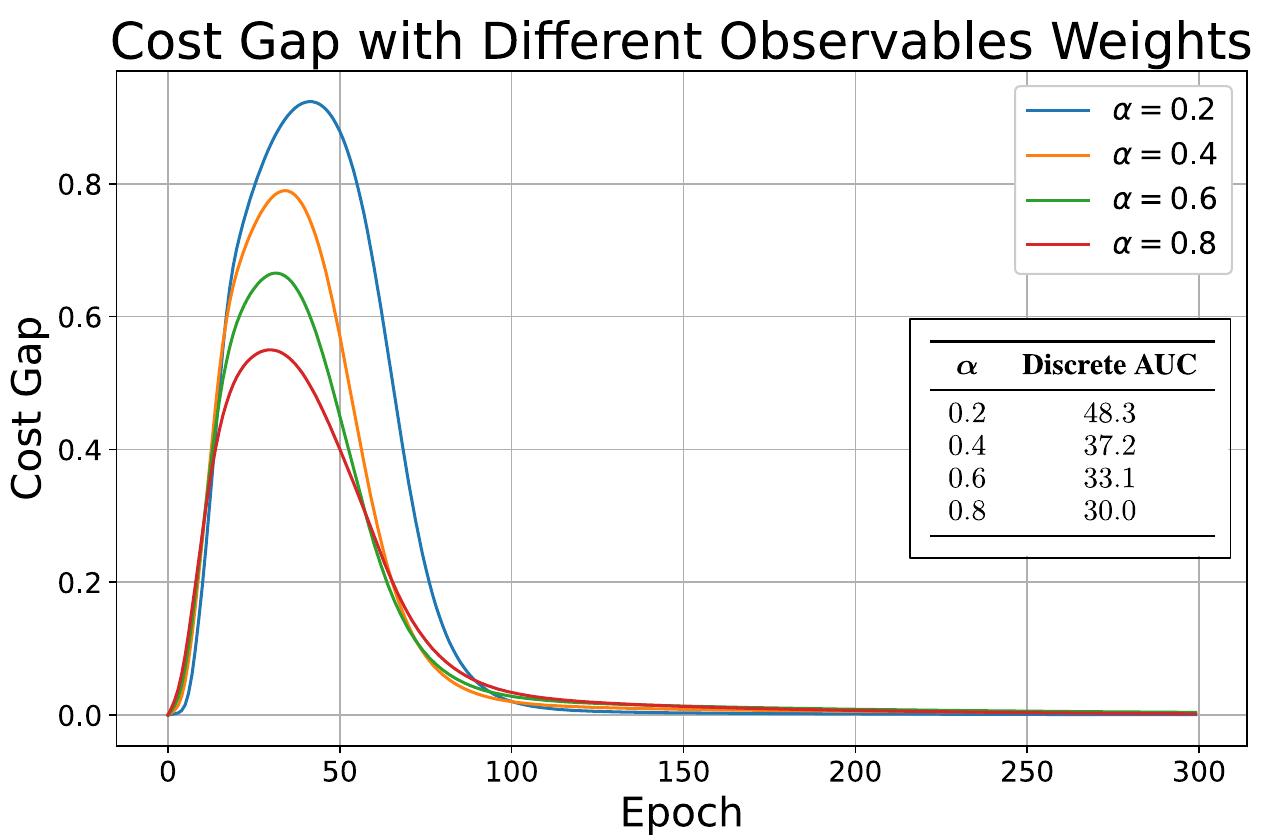}
  \caption{The cost function value gap curves and corresponding discrete AUC between WA-QNG and QNG under different values of $\alpha$ for the Hamiltonian $H=Z_1Z_2+Z_2Z_3+\alpha X_1+(3-2\alpha)X_2+ \alpha X_3$. A smaller $\alpha$ increases the weight of the subsystem on the second qubit in the output, leading to a scenario where WA-QNG should outperform QNG more significantly. This figure indicates that the theoretical analysis aligns with the numerical results that WA-QNG indeed achieves greater performance improvement over QNG as $\alpha$ gradually decreases.}
  \label{fig:cost_gap}
\end{figure*}

To intuitively quantify the performance gap between WA-QNG and standard QNG, we use the difference in cost function values at the same optimization step on the learning curves as an indicator. For a fair comparison, this difference is normalized by dividing it by the difference between the initial and converged cost function value. The cost function value gap curve during training and the discrete area under the gap curve (also representing the discrete area between the learning curves of WA-QNG and QNG, namely the AUC) are presented in \cref{fig:cost_gap}.

As $\alpha$ decreases from $0.8$ to $0.2$, the contribution of the subsystem on the second qubit increases. The numerical results align with the theoretical analysis, as the discrete AUC indeed increases with decreasing $\alpha$. It indicates that capturing the different weights and contributions of each subsystem in WA-QNG effectively improves optimization performance compared to standard QNG. This result implies that WA-QNG is more suitable for situations where the coefficients of each observable term in the $k$-local Hamiltonian vary significantly and are unbalanced. 

\subsection{Locality of Hamiltonian} \label{NQ}
As mentioned in \cref{LQ}, when the entire system becomes significantly larger than the subsystems that directly contribute to the output, the sensitivity of each parameter in the total system differs considerably from that of each subsystem. Under this condition, WA-QNG is expected to outperform standard QNG to a greater extent. To gain a clearer understanding, we conduct numerics using an $n$-qubit toy Ising model Hamiltonian, with $n$ varying from $2$ to $5$, while fixing the maximum locality of the Hamiltonian terms to be $2$:
\begin{equation}
    H = \sum _{i=1}^{n-1}Z_i Z_{i+1} + \sum_{i=1}^{n}X_i
    \label{qubits}
\end{equation}

\begin{figure*}[ht]
  \centering
  \includegraphics[width=0.45\textwidth]{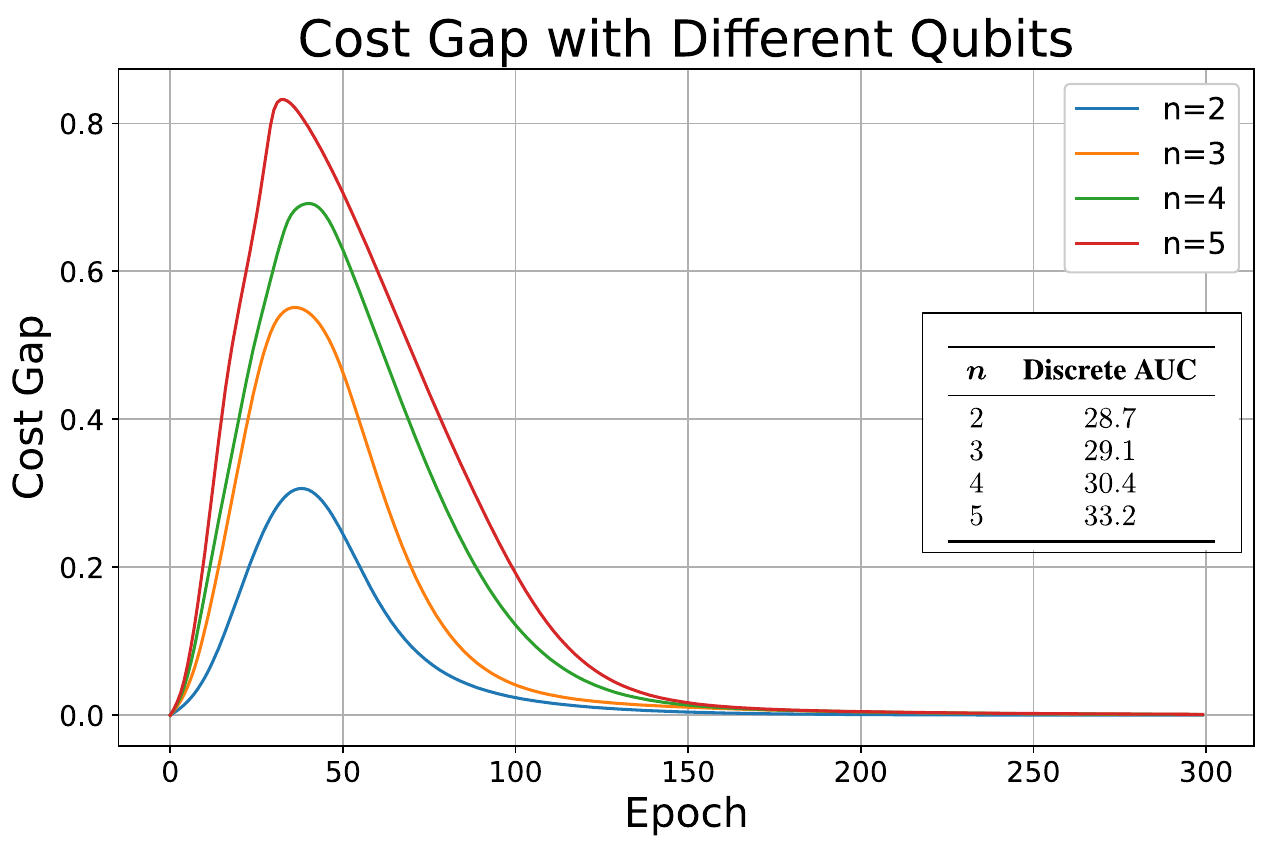}
  \caption{The cost function value gap curves and corresponding discrete AUC between WA-QNG and QNG under different values of $n$ for the Hamiltonian $H = \sum _{i=1}^{n-1}Z_i Z_{i+1} + \sum_{i=1}^{n}X_i$. A larger $n$ decreases the ratio between the locality of Hamiltonian term and the global system size becomes smaller, and increases the difference in parameter sensitivity of each subsystem compared to that of the entire system, leading to a scenario where WA-QNG should outperform QNG more significantly. This figure shows that the theoretical analysis aligns with the numerical results, indicating that WA-QNG achieves greater performance improvement over QNG as $n$ increases with a fixed locality factor $k$.}
  \label{fig:cost_qubit}
\end{figure*}

As $n$ increases, the ratio between the locality of Hamiltonian term and the global system size becomes smaller, and the sensitivity of each parameter in each subsystem differs more significantly from that of the entire system, as each observable term in $H$ is at most $2$-local. Consequently, WA-QNG is expected to perform increasingly better as $n$ increases compared to standard QNG. Similar to the previous subsection, the cost function value gap curve during training and the discrete AUC are shown in \cref{fig:cost_qubit}. For a fair comparison, this difference is also normalized by dividing it by the difference between the initial and converged cost function value.

The numerical results agree with the theoretical analysis. As $n$ increases, the discrete AUC also increases, indicating a more significant performance improvement for WA-QNG. This suggests that WA-QNG is particularly well-suited for scenarios where the total system size $n$ is much larger than the locality factor $k$ for a $k$-local Hamiltonian. QNG is a special case of WA-QNG when $k=n$.

\section{Discussion and Conclusion}\label{con}
In this work, we mainly introduce the WA-QNG, which accounts for the different weights and contributions of each subsystem in the optimization process. In particular, we propose using the matrix $W=\frac{2}{\sum_mh_m^2}\sum_mh_m^2T_m$ instead of the quantum Fisher information matrix of the entire system in each optimization step. We provide three perspectives to explain the effectiveness and potential advantages of WA-QNG. Firstly, the matrix $W$ in WA-QNG can be interpreted as an approximation of the weighted average of the quantum Fisher information matrix of each subsystem contributing to the output. Secondly, from an optimization view, we illustrate that WA-QNG can be derived from a constrained optimization problem where the Euclidean distance in the parameter space is replaced by a weighted sum over the 2-norm distances between density matrices. We further explain that WA-QNG can also be derived from an information geometric perspective, where it emerges as a pullback metric. Additionally, we demonstrate that the optimization task can be approximately transformed into a non-linear least squares problem, where WA-QNG is equivalent to the Gauss-Newton method.

To evaluate the performance of WA-QNG, we conduct numerical simulations on the variational eigensolver for the Ising model and Heisenberg model. The results indicate that WA-QNG achieves superior optimization performance compared to standard QNG. Additionally, we perform further numerics to investigate the source of WA-QNG’s advantage. Our findings indicate two key factors. The first is accounting for the weights of each subsystem indeed improves optimization performance. The second is the Hilbert-Schmidt metric tensor for each subsystem provides a better representation of parameter sensitivity within subsystems compared to using the quantum Fisher information matrix of the entire system. The numerical results are consistent with the theoretical analysis.

In \cref{OI} and \cref{WI}, we mention that WA-QNG has a stronger theoretical explanation from both its approximation to quantum Fisher information matrix and its connection with Gauss-Newton method when each subsystem state is close to being pure and does not change significantly with parameters. However, this assumption is not necessary for WA-QNG to perform better than standard QNG. In the numerics, we observe that WA-QNG still outperforms standard QNG, even when the final target is an entanglement state where each subsystem is mixed.

Since the focus of this paper is to introduce the novelty of WA-QNG itself, we track the exact expectation values and metric tensors of the quantum circuit. In practical applications, these quantities can only be estimated through finite shots, which introduces shot noise. Additionally, we assume that the circuit is noise-free, which is not the case in real-world implementations. Investigating the effects of shot noise and circuit noise on WA-QNG is left as a potential direction for future work.

This work primarily focuses on optimization for the quantum eigensolver. However, since QNG can also be applied to optimize other variational quantum algorithms such as the variational quantum classifier in the field of machine learning, we believe WA-QNG is expected to be extendable to these tasks similarly. Given the formulation of WA-QNG, any variational algorithm employing a cost function defined by the expected value of a $k$-local Hamiltonian could potentially benefit from WA-QNG beyond ground state energy problems. While evaluating the performance of WA-QNG in these broader applications is beyond the scope of this work, it remains a promising direction for future research. 

In conclusion, WA-QNG offers a promising and efficient optimization method for variational quantum eigensolvers. By accounting for the weights of each subsystem that contributes to the output, WA-QNG presents a potential research direction for optimization in variational quantum algorithms.

\section*{Acknowledgment}
All authors acknowledge the support from the Dutch National Growth Fund (NGF), as part of the Quantum Delta NL programme. V.D. acknowledges support from the Dutch Research Council (NWO/OCW), as part of the Quantum Software Consortium programme (project number 024.003.03). This project was also co-funded by the European Union (ERC CoG, BeMAIQuantum, 101124342).

\section*{Author Contributions}
This project was conceived by C.S. and H.W.. C.S. and H.W. formulated the theoretical part. V.D. helped C.S. for the design of numerical simulations. C.S. conducted numerical simulations and analyzed the results with assistance from V.D. and H.W. All authors reviewed the paper on both theoretical and numerical parts.

\section*{Data Availability}
The data that support the findings of this study are available upon reasonable request from the authors.

\onecolumn
\clearpage
\appendix
\section*{Appendix}
\section{QNG as a special case of WA-QNG} \label{DG}
When a term $H_m$ in the Hamiltonian acts globally on the entire quantum system, the corresponding nontrivial subsystem $\rho_m$ becomes the full system $\rho$. In this scenario, each $T_m$ no longer depends on the index $m$, causing the terms $\sum_mh_m^2$ in the weighted summation to cancels out with that in the prefactor. As a consequence, the matrix $W$ exactly reduces to the Hilbert-Schmidt metric tensor of the whole quantum system $\rho_{\theta}$ with a constant factor 2, where the $i$-th row and $j$-th column element of the matrix $W$ is as follows:

\begin{equation}
W_{ij}=2\tr(\partial_i\rho_{\theta}\partial_j\rho_{\theta})
\label{eq:wmatrix}
\end{equation}

Now we prove the matrix $W$ is equal to the quantum Fisher information matrix $F$. The variational state $\rho_{\theta}$ on the whole system is a pure state, so we can write the state $\rho_{\theta}$ as $\rho_{\theta}=\ket{\phi}\bra{\phi}$. So we only have to prove the right side of \cref{eq:wmatrix} is equal to that of \cref{eq:fisher}:

\begin{align}
    W_{ij} &=  2\tr(\partial_i \rho \partial_j \rho) \notag \\
    &= 2\tr(\partial_i (\ket{\phi}\bra{\phi}) \partial_j (\ket{\phi}\bra{\phi})) \notag \\
    &= 2\tr((\ket{\partial_i \phi}\bra{\phi} + \ket{\phi}\bra{\partial_i \phi}) ( \ket{\partial_j \phi}\bra{\phi}+\ket{\phi}\bra{\partial_j \phi})) \notag \\
    &= 2\tr(\ket{\partial_i \phi}\braket{\phi |\partial_j \phi}\bra{\phi}+\ket{\partial_i \phi}\bra{\partial_j\phi}+\ket{\phi}\braket{\partial_i \phi|\partial_j \phi}\bra{\phi}+\ket{\phi}\braket{\partial_i \phi|\phi}\bra{\partial_j \phi}) \notag \\
    &= 2\braket{\phi |\partial_j \phi} \braket{\phi | \partial_i \phi} + 2\braket{\partial_i \phi|\phi}\braket{\partial_j \phi | \phi}+2\braket{\partial_j \phi|\partial_i \phi} + 2\braket{\partial_i \phi | \partial_j \phi} \notag \\
    &= 2\braket{\partial_j \phi|\partial_i \phi} + 2\braket{\partial_i \phi | \partial_j \phi} - 2\braket{\phi |\partial_j \phi} \braket{\partial_i \phi | \phi} - 2\braket{\phi | \partial_i\phi}\braket{\partial_j \phi|\phi} \notag \\
    &= 4\operatorname{Re}(\braket{\partial_i \phi|\partial_j \phi}- \braket{\partial_i \phi|\phi}\braket{\phi|\partial_j \phi}) \notag \\
    &= F_{ij}
\end{align}

\section{Estimate Hilbert-Schmidt Metric Tensor via Classical Shadows} 
In this section, we demonstrate that the Hilbert-Schmidt metric tensor $T$ \footnote{Here for notation simplify, we omit the subscript for each $T_m$ in the definition of the matrix $W$ in \cref{MF}} used in WA-QNG can be efficiently estimated by classical shadow and bound the shots required to obtain the element $T_{ij}$. To avoid confusion in the derivation, we will use calligraphic font $\mathcal{T}$ to represent the Hilbert-Schmidt metric tensor in the following section.
\label{CS}
\subsection{Parameter-Shift Rule} \label{psr}
 First, the parameter-shift rule can be applied to compute each element of the matrix $\mathcal{T}$:
\begin{align}
    \mathcal{T}_{ij} &= 2\tr(\partial_i \rho_{\theta} \partial_j \rho_{\theta}) \notag \\
    &= \frac{1}{2}\tr \Big( (\rho_{\theta+\frac{\pi}{2}e_i}-\rho_{\theta-\frac{\pi}{2}e_i})(\rho_{\theta+\frac{\pi}{2}e_j}-\rho_{\theta-\frac{\pi}{2}e_j})\Big) \notag \\
    &=\frac{1}{2}\Big(\tr(\rho_{\theta+\frac{\pi}{2}e_i}\rho_{\theta+\frac{\pi}{2}e_j})-\tr(\rho_{\theta+\frac{\pi}{2}e_i}\rho_{\theta-\frac{\pi}{2}e_j})-\tr(\rho_{\theta-\frac{\pi}{2}e_i}\rho_{\theta+\frac{\pi}{2}e_j})+\tr(\rho_{\theta-\frac{\pi}{2}e_i}\rho_{\theta-\frac{\pi}{2}e_j})\Big)
\label{eq: para-shift}
\end{align}

where $e_i$ represents the unit vector with the $i$-th element set to one and all other elements set to zero. To estimate $\mathcal{T}_{ij}$, we only need to estimate the four terms in \cref{eq: para-shift} respectively. To estimate the entire matrix $\mathcal{T}$, we can estimate each element individually, meaning the total cost scales quadratically with the number of parameters. Thus, if we can bound the cost of estimating the term like $\tr(\rho\sigma)$, we can also bound the total cost. In our case, where the Hamiltonian is $k$-local, we show the cost of estimating such term $\tr(\rho\sigma)$ via classical shadow is exponential to the subsystem size $k$ rather than the size of the whole system $n$. 

\subsection{Classical Shadow}
The classical shadow technique constructs a series of unbiased estimators $\hat{\rho}^{(t)}$ ($1 \leq t \leq T$, where $T$ is the total number of the classical shadows constructed) for a state $\rho$, with the property that $\operatorname{E}[\hat{\rho}^{(t)}]=\rho$. Each $\hat{\rho}^{(t)}$ is represented as:
\begin{equation}
\hat{\rho}^{(t)}=\bigotimes_{i=1}^{n}(3U_i^{\dag}\ket{\hat{b}_i}\bra{\hat{b}_i}U_i-\mathbb{I})
\label{eq:cs-data}
\end{equation}

where $n$ is the system size, $b$ is a binary string obtained by measurements, and $b_i$ represents the $i$-th bit of $b$ (either $0$ or $1$). $U$ denotes the corresponding random Pauli gate applied to the $i$-th qubit. For more details on the data acquisition process in the classical shadow technique, please refer to \cite{huang2020predicting} and \cite{sack2022avoiding}. Two important properties of each estimator $\hat{\rho}^{(t)}$ are as fellows:

\begin{equation}
\operatorname{E}[\tr(\hat{\rho}^{(t)}O)]= \tr(\rho O)
\label{eq:cs-e}
\end{equation}

\begin{equation}
\operatorname{Var}[\tr(\hat{\rho}^{(t)}O)] \leq 2^{w(O)}\tr(O^2)
\label{eq:cs-v}
\end{equation}

where $w(O)$ represents the number of qubits on which the observable $O$ acts nontrivially. For the details of derivation of \cref{eq:cs-e} and \cref{eq:cs-v}, please refer to the paper \cite{neven2021symmetry}.

To reduce error, an empirical average is taken over all samples to construct the estimator $\hat{\rho}$:
\begin{equation}
\hat{\rho} = \frac{1}{T}\sum_i^T \hat{\rho}^{(t)}
\label{eq:cs-average}
\end{equation}

From \cref{eq:cs-e} and \cref{eq:cs-v}, the following properties of the estimator $\hat{\rho}$ can be derived:

\begin{equation}
\operatorname{E}[\tr(\hat{\rho}O)]= \tr(\rho O)
\label{eq:cs-ae}
\end{equation}

\begin{equation}
\operatorname{Var}[\tr(\hat{\rho}O)] \leq \frac{2^{w(O)}\tr(O^2)}{T}
\label{eq:cs-av}
\end{equation}

\subsection{Construct Estimator for Hilbert-Schmidt Metric Tensor}
As discussed in \cref{psr}, estimating the Hilbert-Schmidt metric tensor via the classical shadow technique requires constructing an estimator for terms like $\tr(\rho \sigma)$.  Similar to the estimator used for estimating purity in \cite{huang2020predicting} and \cite{neven2021symmetry}, the following estimator can be constructed for the term like $\tr(\rho \sigma)$. For simplicity, we denote $p=\tr(\rho \sigma)$, then the corresponding estimator $\hat{p}$ is:

\begin{equation}
\hat{p} = \frac{1}{T^2}\sum_{ij} \tr(\hat{\rho}^{(i)}\hat{\sigma}^{(j)})
\label{eq:estimator}
\end{equation}

where each $\hat{\rho}^{(i)}$ and $\hat{\sigma}^{(j)}$ ($ 1 \leq i,j \leq T$) is obtained using the classical shadow technique as described in \cref{eq:cs-data}. Because $\hat{\rho}^{(i)}$ and $\hat{\sigma}^{(j)}$ are independent, we have:
\begin{align}
    \operatorname{E}[\hat{p}] &= \operatorname{E}[\frac{1}{T^2}\sum_{ij} \tr(\hat{\rho}^{(i)}\hat{\sigma}^{(j)})] \notag \\
    &= \frac{1}{T^2}\sum_{ij} \operatorname{E}[\tr(\hat{\rho}^{(i)}\hat{\sigma}^{(j)})] \notag \\
    &= \frac{1}{T^2}\sum_{ij} \tr(\operatorname{E}[\hat{\rho}^{(i)}]\operatorname{E}[\hat{\sigma}^{(j)}]) \notag \\
    &= \tr(\rho \sigma) \notag \\
    &= p
\end{align}

Hence, the estimator $\hat{p}$ is also an unbiased estimator for $p$. To bound the computational cost, we also need to bound the variance of the estimator $\hat{p}$.

\subsection{Bounding Variance}
According to the definition of the variance of a random variable, we have:
$\hat{p}$.
\begin{align}
    \operatorname{Var}[\hat{p}] &= \operatorname{E}[(\hat{p}-p)^2] \notag \\
    &= \operatorname{E}\Big[\Big(\frac{1}{T^2}\sum_{ij}(\tr(\hat{\rho}^{(i)}\hat{\sigma}^{(j)})-\tr(\rho \sigma))    \Big)^2\Big] \notag \\
    &= \frac{1}{T^4} \sum_{ij}\sum_{kl} \operatorname{E}\Big[\big(\tr(\hat{\rho}^{(i)}\hat{\sigma}^{(k)})-\tr(\rho\sigma)\big)\big(\tr(\hat{\rho}^{(j)}\hat{\sigma}^{(l)})-\tr(\rho\sigma)\big)\Big]
    \label{eq:variance}
\end{align}

The summation in \cref{eq:variance} can be divided into the following three cases. Suppose the density operator $\rho$ and $\sigma$ are systems of $n$-qubit.

1. When $i\neq k$ and $j\neq l$: There are $T^2(T-1)^2$ terms. For each term, we have:
\begin{align}
    &\operatorname{E}\Big[\big(\tr(\hat{\rho}^{(i)}\hat{\sigma}^{(j)}-\tr(\rho\sigma)\big)\big(\tr(\hat{\rho}^{(k)}\hat{\sigma}^{(l)})-\tr(\rho\sigma)\big)\Big] \notag \\
    &= \big(\operatorname{E}[\tr(\hat{\rho}^{(i)}\hat{\sigma}^{(j)})]-\tr(\rho\sigma)\big)\big(\operatorname{E}[\tr(\hat{\rho}^{(j)}\hat{\sigma}^{(k)})]-\tr(\rho\sigma)\big) \notag \\
    &= 0
    \label{eq:variance1}
\end{align}

2. When $i=k$ but $j\neq l$, or $j=l$ but $i\neq k$: There are $2T^2(T-1)$ terms. Without loss of generality, we take the case where $i=k$ but $j\neq l$ as an example. The calculation in another case is the same. For each term, we have:
\begin{align}
    &\operatorname{E}\Big[\big(\tr(\hat{\rho}^{(i)}\hat{\sigma}^{(j)})-\tr(\rho\sigma)\big)\big(\tr(\hat{\rho}^{(i)}\hat{\sigma}^{(l)})-\tr(\rho\sigma)\big)\Big] \notag \\
    &=\operatorname{E}\big[\tr(\hat{\rho}^{(i)}\hat{\sigma}^{(j)})\tr(\hat{\rho}^{(i)}\hat{\sigma}^{(l)})\big] -\tr^2(\rho \sigma) \notag \\
    &=\operatorname{E}\big[\tr(\hat{\rho}^{(i)\otimes 2} \hat{\sigma}^{(j)}\otimes \hat{\sigma}^{(l)} )\big]-\tr^2(\rho\sigma) \notag\\
    &=\operatorname{E}\big[\tr^2(\hat{\rho}^{(i)}\sigma)\big] - \tr^2(\rho\sigma) \notag \\
    &=\operatorname{Var}[\tr(\hat{\rho}^{(i)}\sigma))] \notag \\
    &\leq 2^{w(\sigma)}\tr(\sigma^2) \notag \\
    &\leq 2^n
    \label{eq:variance2}
\end{align}

The third equality in \cref{eq:variance2} relies on the property that, when the unbiased estimators $\hat{\rho}$ and $\hat{\sigma}$ are independent, then $\operatorname{E}[\hat{\rho}\otimes \hat{\sigma}]=\rho \otimes \sigma$. The details of this property can be found in \cite{neven2021symmetry}. The first inequality is from \cref{eq:cs-v}. 

3. When $i=k$ and $j=l$: There are $T^2$ terms. For each term, we have:
\begin{align}
    &\operatorname{E}\Big[\big(\tr(\hat{\rho}^{(i)}\hat{\sigma}^{(j)})-\tr(\rho\sigma)\big)\big(\tr(\hat{\rho}^{(i)}\hat{\sigma}^{(j)})-\tr(\rho\sigma)\big)\Big] \notag \\
    &=\operatorname{E}\big[\tr^2(\hat{\rho}^{(i)}\hat{\sigma}^{(j)})\big]-\tr^2(\rho \sigma) \notag \\
    &=\operatorname{Var}[\tr(\hat{\rho}^{(i)}\hat{\sigma}^{(j)}))] \notag \\
    &=\operatorname{Var}[\tr(S \hat{\rho}^{(i)}\otimes \hat{\sigma}^{(j)}))] \notag\\
    &\leq 2^{w(S)} \tr(S^2) \notag\\
    &= 2^{4n}
\end{align}

The third equality follows from the property $\tr(S\rho \otimes\sigma)=\tr(\rho \sigma)$, where $S\in \mathbb{C}^{2^{2n}\times2^{2n}}$ is the \textit{SWAP} operator. The details of this property can be found in \cite{mele2024introduction}. The inequality arises from \cref{eq:cs-v}. The last equality holds because $S$ acts on $2^{2n}$ qubits and satisfies $S^2=I$.

Hence, the final result of \cref{eq:variance} shall be upper bounded by:
\begin{align}
    \operatorname{Var}[\hat{p}] &=\frac{1}{T^4} \sum_{ij}\sum_{kl} \operatorname{E}\Big[\big(\tr(\hat{\rho}^{(i)}\hat{\sigma}^{(k)})-\tr(\rho\sigma)\big)\big(\tr(\hat{\rho}^{(j)}\hat{\sigma}^{(l)})-\tr(\rho\sigma)\big)\Big]\notag \\
    &\leq \frac{1}{T^4}\big(2T^2(T-1)2^n+T^22^{4n}\big)\notag \\
    &\leq \frac{2^{n+1}}{T}+\frac{2^{4n}}{T^2}
    \label{eq:variance-bound}
\end{align}

\subsection{Bounding Shots}
Using the bound of variance \cref{eq:variance-bound}, we can derive the upper bound of total shots required in the estimation for the term like $\tr(\rho\sigma)$. Following the assumption in the reference paper \cite{sack2022avoiding} that, the $T$ can be very large where the expression of \cref{eq:variance-bound} is dominated by the first term. Following Chebyshev's inequality:
\begin{equation}
    \operatorname{Pr}[|\hat{p}-p|\geq \epsilon] \lesssim \frac{2^{n+1}}{T\epsilon^2}
\end{equation}

Then, a measurement budget that scales as
\begin{equation}
    T \geq \frac{2^{n+1}}{\epsilon^2\delta}
\end{equation}
with probability $1-\delta$ suffice to control the estimation error below $\epsilon$. Hence, the lower bound of shots required to estimate a term like $\tr(\rho\sigma)$ is $\operatorname{O}(2^{n+1})$, where $n$ is the system size of the quantum density operator $\rho$ and $\sigma$.

In our case, there are four terms like $\tr(\rho\sigma)$ in the \cref{eq: para-shift} required to estimate for the element $\mathcal{T}_{ij}$, and each term is only $k$-local. Hence the cost of shots required to estimate each element $\mathcal{T}_{ij}$ is $\operatorname{O}(4\cdot2^{k+1})$, which is exponential to the subsystem size $k$ instead of the entire system size $n$.

\section{Approximate Quantum Fisher Information via Hilbert-Schmidt Metric Tensor} \label{AQFI}
In this section, we provide a simple proof that the Hilbert-Schmidt metric tensor serves as an approximation to the quantum Fisher information matrix $F$ when the state $\rho$ is close to being pure and does not change significantly with parameters. Moreover, this approximation becomes exact when $\rho$ is pure.

For a state $\rho=\sum_{k=1}^n r_k\ket{r_k}\bra{r_k}$, suppose the dominant eigenvector \cite{koczor2022quantum1,koczor2022quantum2} is $\ket{r_d}$ with eigenvalue $r_d$. When state $\rho$ is close to being pure, the dominant eigenvalue satisfies $r_d \approx 1$, while all other eigenvalues satisfy $r_k \approx 0$. For the state $\rho$, we show its Hilbert-Schmidt metric tensor can be computed and simplified as:
\begin{align}
    \tr(\partial_i\rho\partial_j\rho)&=\sum_k\frac{\partial r_k}{\partial \theta_i}\cdot\frac{\partial r_k}{\partial \theta_j}+\sum_k r_k^2 \frac{F_k}{2} - \sum_{kl,k\neq l} r_kr_l\cdot2\operatorname{Re}[\braket{\partial_i r_k|r_l}\braket{r_k|\partial_jr_l}]\notag \\
    &\approx \frac{r_d^2}{2}(F_d)_{ij} \notag \\
    &\approx \frac{1}{2}F_{ij}
    \label{approxi}
\end{align}

where $(F_d)_{ij}$ represents the $(i,j)$-th element of the quantum Fisher information matrix of the dominant state. The derivation of the first equation will be explained in detail in the following discussion. When the state $\rho$ is close to being pure, the other eigenvalues are higher-order infinitesimals compared to the dominant eigenvalues. Consequently, the second and third terms in the first equation can be approximated as $\frac{r_d^2}{2}(F_d)_{ij}$. If the state does not change drastically with respect to the parameters, the first term will also be small. In practical applications, this can be achieved by initializing the variational quantum circuit with low entanglement and setting the learning rate to a small value. Moreover, because $r_d \approx 1$ and the state $\ket{r_d}$ dominants $\rho$, we can obtain $\frac{r_d^2}{2}(F_d)_{ij} \approx \frac{1}{2}F_{ij}$. However, this approximation assumption is not necessary for WA-QNG to be well defined or to outperform standard QNG.

Now, we provide a brief derivation of the first equation. For simplicity of notation, we denote the three terms in $\partial_i \rho = \sum_k \partial_ir_k\ket{r_k}\bra{r_k}+\sum_kr_k\ket{\partial_ir_k}\bra{r_k}+\sum_kr_k\ket{r_k}\bra{\partial_ir_k}$ as $A_i$, $B_i$, and $C_i$ respectively. Similarly, for $\partial_j\rho$ we can also denote $A_j$, $B_j$ and $C_j$ for the three terms analogously. Hence, we can express:
\begin{align}
    &\tr(\partial_i\rho\partial_j\rho) \\
    &= \tr(A_iA_j)+\tr(A_iB_j)+\tr(A_iC_j)+\tr(B_iA_j)+\tr(B_iB_j)+\tr(B_iC_j)+\tr(C_iA_j)+\tr(C_iB_j)+\tr(C_iC_j) \notag
    \label{ABC}
\end{align}

We can compute these terms separately.

1. $\tr(A_iA_j)$:
\begin{align}
    \tr(A_iA_j) &= \tr\big(\sum_k \frac{\partial r_k}{\partial \theta_i}\frac{\partial r_k}{\partial \theta_j}\ket{r_k}\bra{r_k} \big)\notag \\
    &= \sum_k \frac{\partial r_k}{\partial \theta_i}\frac{\partial r_k}{\partial \theta_j}
\end{align}

2. $\tr(A_iB_j)$ and $\tr(A_iC_j)$:
\begin{align}
    \tr(A_iC_j)+\tr(A_iB_j) &= \sum_k \frac{\partial r_k}{\partial \theta_i}r_k \braket{\partial_jr_k|r_k} + \sum_k \frac{\partial r_k}{\partial \theta_i}r_k \braket{r_k|\partial_jr_k} \notag \\
    &= 0
\end{align}

3. $\tr(B_iC_j)$ and $\tr(C_iB_j)$:
\begin{align}
    \tr(B_iC_j)+\tr(C_iB_j) &= \sum_kr_k^2\braket{\partial_j r_k|\partial_i r_k}+\sum_kr_k^2\braket{\partial_i r_k|\partial_j r_k} \notag \\
    &=  \sum_k2r_k^2 \operatorname{Re}[\braket{\partial_i r_k|\partial_j r_k}]
\end{align}

4. $\tr(B_iB_j)$ and $\tr(C_iC_j)$:
\begin{align}
    \tr(B_iB_j)+\tr(C_iC_j) &= \sum_{kl} r_kr_l\big(\braket{r_k|\partial_j r_l}\braket{r_l|\partial_ir_k}+\braket{\partial_i r_k|r_l}\braket{\partial_i r_k|r_l}\braket{\partial_j r_l|r_k}\big) \notag \\
    &= -\sum_k r_k^2\cdot2\operatorname{Re}[\braket{\partial_ir_k|r_k}\braket{r_k|\partial_jr_k}]-\sum_{kl,k\neq l} r_kr_l\cdot2\operatorname{Re}[\braket{\partial_ir_k|r_l}\braket{r_k|\partial_jr_l}]
\end{align}

5. $\tr(C_iA_j)$ and $\tr(B_iA_j)$:
\begin{align}
    \tr(C_iA_j)+\tr(B_iA_j) &= r_k^2\braket{r_k|\partial_ir_k}+r_k^2\braket{\partial_i r_k|r_k} \notag \\
    &= 0
\end{align}

By combining these results, we obtain the first equation in \cref{approxi}.
\section{Deriving WA-QNG Optimization Step from Geometric Interpretation} \label{GEO}
According to the Lagrange multiplier method, the constrained optimization problem in \cref{waqng_problem} can be formulated as:
\begin{equation}
    d^{*}=\argmin_{d} f(\theta + d) + \lambda \Big( \frac{2}{\sum_mh_m^2}\sum_m h_m^2 \Vert\rho_m(\theta+d)-\rho_m(\theta) \Vert_2^2
    -\epsilon \Big)
    \label{lag1}
\end{equation}

For the trace $2$-norm term, applying the first-order Taylor expansion to $\rho_m(\theta+d)$, we obtain:
\begin{align}
    \Vert\rho_m(\theta+d)-\rho_m(\theta) \Vert_2^2 &\approx \Vert \rho_m(\theta) + \sum_i \partial_i\rho_m(\theta) d_i-\rho_m(\theta) \Vert_2^2 \notag \\
    &=\Vert \sum_i\partial_i\rho_m(\theta) d_i \Vert_2^2 \notag \\
    &= \tr\Big(\sum_i\sum_j \partial_i\rho_m \partial_j\rho_m  d_i d_j\Big) \notag \\
    &= \sum_i \sum_j \tr(\partial_i\rho_m\partial_j \rho_m) d_i d_j 
    \label{tylor}
\end{align}
    
Substituting \cref{tylor} into \cref{lag1} and applying the first-order Taylor expansion to $f(\theta+d)$, we obtain:
\begin{align}
    d^{*} &\approx \argmin_{d} f(\theta) + \nabla f(\theta)^{T}d + \frac{2\lambda}{\sum_mh_m^2}\sum_m h_m^2 \sum_i \sum_j \tr(\partial_i\rho_m\partial_j \rho_m)d_id_j -\lambda\epsilon \notag \\
    &= \argmin_{d} f(\theta) + \nabla f(\theta)^{T}d + \sum_i\sum_j \   \Big(\frac{2\lambda}{\sum_mh_m^2}\sum_m h_m^2\tr(\partial_i\rho_m\partial_j \rho_m) \Big) d_id_j -\lambda\epsilon \notag \\
    &= \argmin_{d} f(\theta) + \nabla f(\theta)^{T}d + \lambda d^{T}Wd -\lambda\epsilon
    \label{lag2}
\end{align}

where the matrix $W$ is exactly the same matrix defined in WA-QNG in \cref{eq:wa-qng}. Since we are computing the minimum,  \cref{lag2} should satisfy the Karush–Kuhn–Tucker (KKT) conditions \cite{kuhn1951proceedings}. Here, it simply means that the derivative of the right side with respect to $d$ should be zero:
\begin{align}
    0 &= \nabla f(\theta) +2\lambda Wd \notag \\
    d & = -\frac{1}{2\lambda} W^{+}\nabla f(\theta)
    \label{lag3}
\end{align}

\cref{lag3} indicates the optimal update direction in WA-QNG. Since the Lagrange multiplier $\lambda$ can be absorbed into the learning rate, the above formula can be exactly transformed into the update formula of WA-QNG, as given in \cref{eq:wa-qng}.

\section{Relation with Gauss-Newton Method Supplement} \label{RGCI}
Here we verify the relation $W=2J_r^TJ_r$. The $(i,j)$-th element of the Jacobian $J_r$ is:
\begin{align}
    (J_r)_{ij} &= \frac{h_i}{\sqrt{\sum_mh_m^2}}\frac{\partial (\operatorname{vec}(\rho_i)-\operatorname{vec}(\tilde{H}_i))}{\partial\theta_j} \notag \\
    &= \frac{h_i}{\sqrt{\sum_mh_m^2}}\frac{\partial\operatorname{vec}(\rho_i)}{\partial \theta_j}
    \label{eq:jac}
\end{align}

Hence, the $(i,j)$-th element of $J_r^TJ_r$ should be:
\begin{align}
    (J_r^TJ_r)_{ij} &= \sum_k (J_r^T)_{ik}(J_r)_{kj} \notag \\
    &= \sum_k\frac{h_k}{\sqrt{\sum_mh_m^2}}\frac{h_k}{\sqrt{\sum_mh_m^2}} \Big(\frac{\partial\operatorname{vec}(\rho_k)}{\partial \theta_i} \cdot\frac{\partial\operatorname{vec}(\rho_k)}{\partial \theta_j}\Big)\notag \\
    &= \frac{1}{\sum_m h_m^2}\sum_kh_k^2\tr(\partial_i\rho_k\partial_j\rho_k) \notag \\
    &= \frac{1}{\sum_mh_m^2}\sum_mh_m^2(T_m)_{ij} \notag \\
    &= \frac{1}{2}W_{ij}
    \label{eq:jac2}
\end{align}
Hence, we have proved the relation $W=2J_r^TJ_r$.

\twocolumn
\bibliographystyle{unsrt}  
\bibliography{references}

\end{document}